\documentclass{ar-1col}

\newcommand{\ebf}[1]{#1}
\newcommand{\ch}[1]{#1}
\newcommand{\ci}[1]{#1}

\usepackage[comma]{natbib}
\usepackage{url}
\usepackage{amsmath}
\usepackage{amssymb}
\usepackage{amsmath,upgreek,bm}
\usepackage{mathtools}
\usepackage{scalerel,stackengine}
\usepackage{ulem}
\usepackage{subcaption}
\stackMath

\makeatletter
\DeclareTextCompositeCommand{\r}{OT1}{A}{%
  \leavevmode\vbox{%
    \offinterlineskip
    \ialign{\hfil##\hfil\cr\char23\cr\noalign{\kern-1.15ex}A\cr}%
  }%
}
\makeatother

\newsavebox{\foobox}
\newcommand{\slantbox}[2][0]{\mbox{%
    \sbox{\foobox}{#2}%
    \hskip\wd\foobox
    \pdfsave
    \pdfsetmatrix{1 0 #1 1}%
    \llap{\usebox{\foobox}}%
    \pdfrestore
}}
\newcommand\unslant[2][-.25]{%
    \mkern1.3mu%
    \ThisStyle{\slantbox[#1]{$\SavedStyle#2$}}%
    \mkern-1.3mu%
}
\newcommand{\utheta}{\boldsymbol{\mathord{\unslant\theta}}}
\newcommand{\ueta}{\boldsymbol{\mathord{\unslant\eta}}}
\newcommand{\uepsilon}{\boldsymbol{\mathord{\unslant\epsilon}}}

\jname{Annu. Rev. Stat. Appl. 2023.}
\jvol{10:623–49}
\jyear{First published as a Review in Advance on
November 22, 2022}
\doi{10.1146/10.1146/annurev-statistics-033021-012225}

\renewcommand*{\vec}[1]{\mathbf{#1}}
\newcommand*{\mat}[1]{\mathbf{#1}}
\DeclareMathOperator{\e}{e}
\DeclareMathOperator{\dd}{d}
\DeclareMathOperator{\RV}{RV}
\begin{document}

\markboth{Hara \& Ford}{Exoplanet detection with radial velocities}

\title{Statistical methods for exoplanet detection with radial velocities} 

\author{Nathan C. Hara$^1$, Eric B. Ford$^{2,3,4,5}$
\affil{$^1$Université de Genève, chemin Pegasi 51, 1290 Versoix, Switzerland; email: nathan.hara@unige.ch}
\affil{$^2$The Pennsylvania State University; email:ebf11@psu.edu}
\affil{$^3$Center for Exoplanets \& Habitable Worlds}
\affil{$^4$Institute for Computational \& Data Sciences}
\affil{$^5$Center for Astrostatistics}
\vspace{2cm}
}

\begin{abstract}
Exoplanets can be detected with various observational techniques. Among them, radial velocity (RV) has the key advantages of revealing the architecture of planetary systems and measuring planetary mass and orbital eccentricities. RV observations are poised to play a key role in the detection and characterization of Earth twins. However, the detection of such small planets is not yet possible due to very complex, temporally correlated instrumental and astrophysical stochastic signals. Furthermore, exploring the large parameter space of RV models exhaustively and efficiently presents difficulties. In this review, we frame RV data analysis as a problem of detection and parameter estimation in unevenly sampled, multivariate time series. The objective of this review is two-fold: to introduce the motivation, methodological challenges, and numerical challenges of RV data analysis to nonspecialists, and to unify the existing advanced approaches in order to identify areas for improvement.
\end{abstract}



\maketitle

\tableofcontents

\section{INTRODUCTION}
\label{sec:intro}

Until recently, our Solar System was the only laboratory in which to study theories of planetary formation and evolution, and to search for life beyond Earth. The presence of planets outside the Solar system, or exoplanets, was  uncertain until the detection of planets orbiting a pulsar \citep{wolszczan1992}. The study of exoplanets as a scientific field accelerated with the detection of 51 Peg b, a 
planet of minimum  $\sim0.5$ Jupiter mass orbiting a Sun-like star in 4.2 days \citep{mayor1995}.



%

The discovery of 51 Peg b, and hundreds of additional exoplanets, was 
based on the radial velocity method, which is the focus of the present work, and relies on the following principle. A star around which a planet revolves has a periodic reflex motion, so it moves back and forth toward the observer with a certain velocity: its radial velocity (RV). By acquiring spectra of a given star at different times, the observer can measure a time series of RV through the Doppler effect. The amplitude of the variations in RV is proportional to the planetary mass, and its shape depends on the orbital eccentricity. 

Mass is one of the most fundamental parameters of a planet.
When the radius is  measured separately via photometry, the combination of mass and radius gives the density, essential to characterizing the internal structure of the planet, 
and the surface gravity, key to interpreting measurements of its atmosphere \citep{kempton2018, batalha2019}. 
The eccentricity is important to understand the formation of planetary systems, especially in multiplanetary systems~\citep[e.g.][]{juric2008}. 

\begin{marginnote}[]
\entry{Radial velocity (RV)}{ Velocity of a given star projected onto the direction observer - star. }
\entry{Stellar spectrum}{Number of photons collected from a given star per wavelength. }
\entry{Spectrograph}{instrument used to
measure the light flux
as a function of its wavelength; can be used to measure the radial velocity of stars}
\end{marginnote}

The RV method plays a key role in understanding the demographics \ch{of planetary systems}, by allowing the detection of planets spanning a much wider range of orbital periods than other detection techniques. 
Moreover, it does not require a precise orientation of the orbital plane, thereby allowing RVs to characterize planetary systems in which the planets have significant mutual inclinations. As such, RV measurements have a unique potential to reveal the architecture of planetary systems with periods up to $\sim 10$ years \citep{ fulton2021, rosenthal2021}



\begin{marginnote}[]
\entry{Photometry}{Number
of photons collected
on a celestial body,
here a star, in a broad
spectral band as a
function of time}
\entry{Transit}{passage of a
planet between a star
and an observer;
results in a periodic
dip in the flux received}
\end{marginnote}



One long-term goal of the \ch{scientific community working on exoplanets} is to detect and characterize the atmospheres of a population of potentially Earth-like planets, and in particular to search for evidence of life.
Characterising the atmosphere of Earth-like planets is beyond the reach of current facilities, but projects \ch{of ground based instruments} such as PCS@ELT \citep{kasper2021}, ANDES@ELT, and direct imaging space mission concepts LUVOIR, HaBex\footnote{\url{https://asd.gsfc.nasa.gov/luvoir/resources/}, \url{https://www.jpl.nasa.gov/habex/documents/}} or nulling interferometer LIFE \citep[ESA, ][]{quanz2021}) \ch{hope to achieve this by 2050}. 

%
%




Measuring the mass of terrestrial planets to \ch{an accuracy of} 20\% or better is essential to interpreting direct imaging and atmospheric observations \citep{batalha2019,crass2021}. Additionally, detecting Earth-like planets with RV measurements prior to the start of a future direct imaging mission would substantially increase its yield. 
The RV technique 
is poised to play a key role in the 
detection and characterisation of Earth-like planets around Sun-like stars, but we do not yet know if it can achieve the required accuracy. 
The Earth produces a velocity variation of the Sun of \ci{$\mathrm{RV}_\oplus$ =} 9 cm/s, \ci{which we use as a standard unit}. The latest-generation of spectrographs (e.g., ESPRESSO (Echelle SPectrograph for Rocky Exoplanets and Stable Spectroscopic Observations), EXPRES (EXtreme PREcision Spectrometer), NEID (NN-EXPLORE Exoplanet Investigations with Doppler spectroscopy)) have demonstrated on-sky precision on a few nights from 3 to 4.5 $\mathrm{RV}_\oplus$ \citep{pepe2021, blackman2020,lin2022}. 
If the measurement noise were independent, then 200 measurements would suffice to measure the mass of an Earth-clone with 20\% precision, but intrinsic stellar variability causes complex, temporally correlated radial velocity signal of the order of at least 5 $\mathrm{RV}_\oplus$. As an example, \ch{the RV of the} Sun \ch{as measured with the HARPS-N spectrograph has a 22 $\mathrm{RV}_\oplus$ standard deviation}.  \ch{Characterizing Earth-like planets requires better methods} for mitigating stellar variability and instrument systematics.
This can be viewed as a particular instance of a general problem: detecting and characterizing periodic signals in multivariate, unevenly sampled time series, corrupted by complex noise.







\ci{The present work overviews} the efforts undertaken to make Earth twins detectable.  The RV data products are presented in \S\ref{sec:data_and_model}. We present an overview of the different challenges in \S\ref{sec:indicators}.
We then present RV analysis in a recursive manner:
Given \ch{summary statistics extracted from raw data and a statistical model of these},  the question of how one detects planets and estimates
their orbital elements is discussed \S\ref{sec:analysis_methods_ts}. 
Given the summary statistics, the question of how
one builds a statistical model of the data is treated in \S\ref{sec:GP_activity}. How to exploit information in lower data products to extract accurate RVs and useful summary statistics is discussed in \S\ref{sec:analysis_methods_dl}.

    \section{DATA AND MODEL}
\label{sec:data_and_model}

\subsection{Doppler shifts}
\begin{figure}[h]
 \includegraphics[width=8cm]{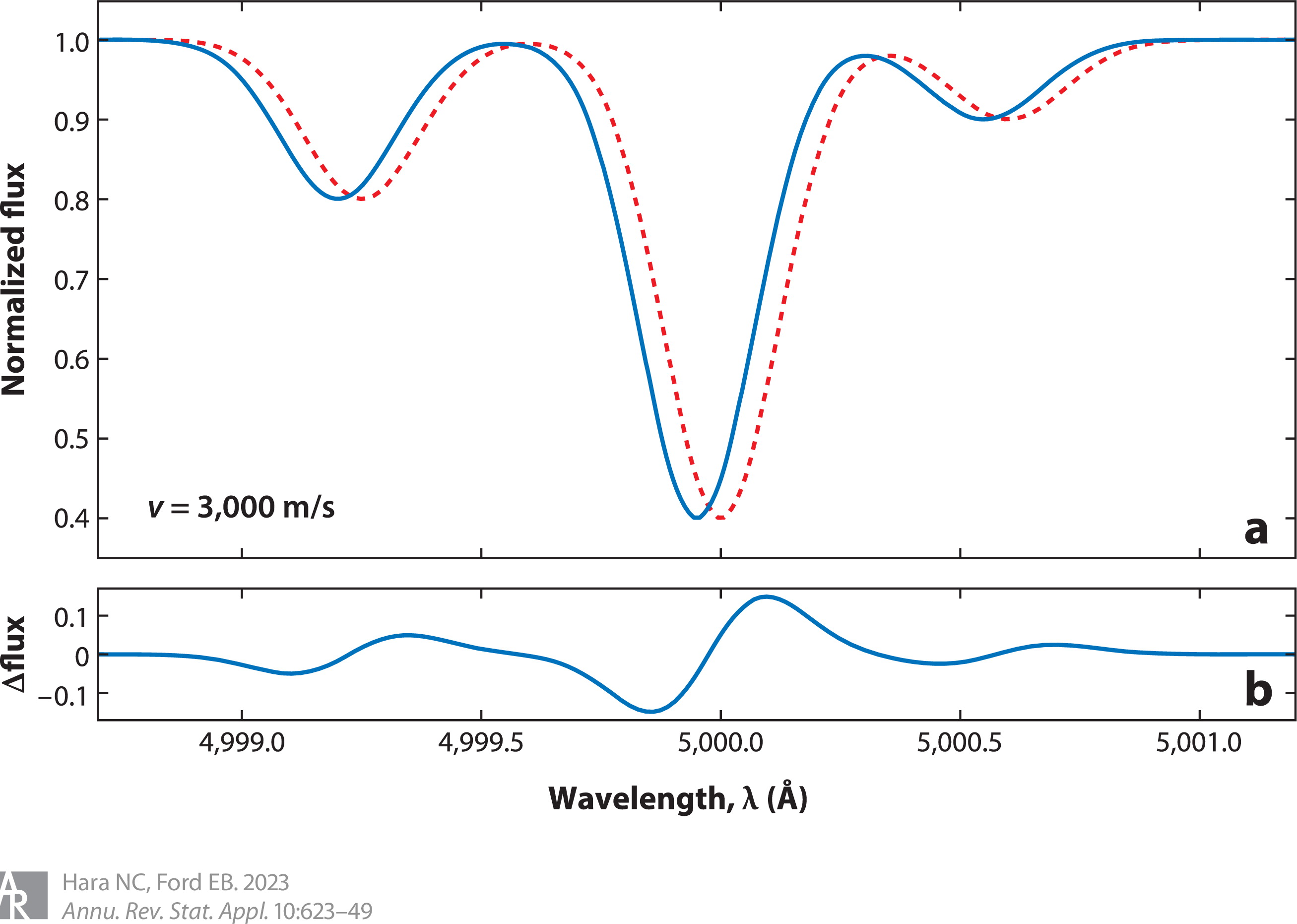}
\caption{(a) Small portion of a spectrum in rest frame (red, dotted) and in observer frame (blue, solid). (b) Difference in
flux between the Doppler-shifted and rest frame spectra. For illustration, we show a large Doppler shift
typical of a binary star system. Abbreviation: v, relative velocity of the stars.}
\label{fig:spectrumchuck}
\end{figure}

\ch{The radial velocity (RV) of a star, its velocity projected onto the line of sight, can be measured due to the Doppler effect}.
If a source emits a photon with wavelength $\lambda_0$, and has a velocity $\vec v$ relative to an observer, then the wavelength of the photon received is given by \begin{equation}
\lambda = \lambda_0 \frac{1 + \frac{1}{c}\vec k \cdot \vec v}{\sqrt{1-\frac{v^2}{c^2}}}  \label{eq:doppler_shift}
\end{equation}
where $\cdot$ is the scalar product and $\vec k$ is the unit vector from the observer to the source \citep{einstein1905}. 
The spectrum of a star contains thousands of absorption lines, short intervals of wavelengths for which photons are absorbed in the upper parts of stellar atmospheres.
As the star moves, the Doppler effect causes the apparent wavelength of spectral lines to change (see Fig.~\ref{fig:spectrumchuck}).
An observer who acquires the spectrum of a given star multiple times can measure changes in the apparent wavelength of each spectral line as a function of time, $t$.  
This maps to measuring $\RV(t)\equiv \vec k \cdot \vec v(t)$ in Eq.~\eqref{eq:doppler_shift} \ci{For more details on the definition of RV, see \cite{lindegren2003} and \cite{lovis2010}.}

Measuring the radial velocity from a spectrum has many steps, including the correction of instrumental effects. 
In this review, we start with 
the estimate of the true stellar spectrum, de-biased from instrumental effects, and transformed to be equivalent to what an observer at the solar system's center of mass would see.
The earlier steps of analysis are briefly presented in Appendix \S\ref{sec:calib}.

\subsection{Forward model of planetary effects}
\label{sec:forward}



\begin{figure}[h]
\includegraphics[width=\linewidth]{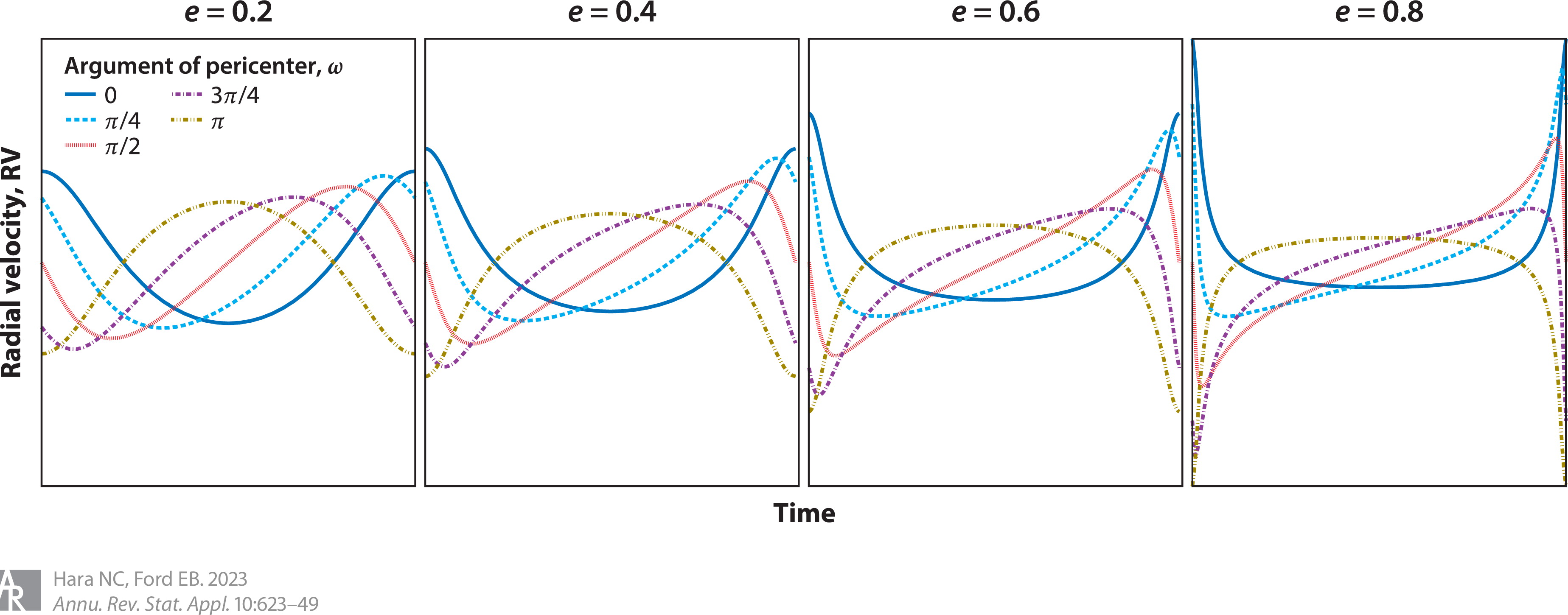}
\caption{Radial velocity (RV) signature of a single planet. Each panel shows the RV signature over one orbital period for a planet with a given
orbital eccentricity e. All curves are computed with the same RV semiamplitude K, and all panels have the same y-axis scale.  }
\label{fig:rv_grid}
\end{figure}

An orbiting planet causes a reflex motion of the star, thus creating RV variations. More precisely, the star and the planet periodically traverse elliptical orbits. The star’s RV projected onto the line of sight, or RV, at time $t$ depends on the planet’s mass and orbital parameters and is given by the so-called Keplerian signal,
\begin{eqnarray}
f(t;K,P,e,\omega,M_0)  & = & K\left[\cos\left(\omega + \nu(t;e,P,M_0)\right) + e \cos \omega\right] \label{eq:rv_planet}
\end{eqnarray}
where $K,P,e,\omega$, and $M_0$ are the velocity amplitude, the orbital period, the orbital eccentricity, the argument of periastron and the mean anomaly describing the motion of the star. The true anomaly, $\nu$, \ch{is an angle that} parametrizes the position of the star \ch{on its orbit}  \citep[see Appendix \S\ref{sec:definitions} and][]{murraycorreia2010}.
%
%
\begin{marginnote}
\entry{Keplerian signal}{RV signal as a function of time resulting from the orbit of a planet, as given in Eq.~\ref{eq:rv_planet} and shown in Fig.~\ref{fig:rv_grid}.}
\end{marginnote}

The eccentricity of the ellipse is confined to $\left[0,1\right)$ for a bound system undergoing periodic motion.  Orbits with $e\simeq~0$ are nearly circular and can be well-approximated by the leading terms of a series expansion in the mean anomaly. Orbits with $e\ge~0.3$ become obviously elongated and the motion becomes noticeably uneven in time, typically leading to a sharp rise and slow fall (or vice versa), as shown in Fig. \ref{fig:rv_grid}.  Very high eccentricities ($0.9<e<1-R_\star/a$, where $R_\star$ is the stellar radius and $a$ is the mean star-planet separation) are very rare, but can create numerical difficulties.
The signal amplitude depends on the mass of the star and the planet, as well as on the orbital eccentricity, according to the formula 
\begin{equation}
    K = \left( \frac{2\pi G}{P} \right)^{\frac{1}{3}} \frac{m \sin i}{(m + M)^{\frac{2}{3}}} \frac{1}{\sqrt{1-e^2}},  \label{eq:semiamp}
\end{equation}
where $M$ is the stellar mass, $m$ is the planet mass,  $G$ the gravitational constant, and $i$ is the angle between plane in which the planet orbits and the plane perpendicular to the line of sight \citep{perryman2011}. \ch{Eq.~\eqref{eq:semiamp} shows that the RV semi-amplitude is proportional to $m/P^{1/3}$: the more massive the star, and the smaller and farther from its star a planet is, the more difficult the planet is to detect. }

\ci{Each observed star has a \ebf{proper motion due to the motion of the star (and the Sun)} around the galactic center.} On the timescale of RV observations, $\sim 1 $ to 20 years, this appears as constant or in some cases a linear trend 
that must be added to Eq.~\eqref{eq:rv_planet} Furthermore,
most Sun-like stars host multiple planets \citep{He2021}.  In principle, the planet-planet interactions cause deviations from the Keplerian model, but for the vast majority of planetary systems, the motion of the star can be well-approximated over the timescale of RV surveys as the linear superposition of the Keplerian orbit due to each planet.
Hence, our nominal physical model for a time series of RV measurements of a star hosting $n$ planets with parameters $ (K_j,P_j,e_j,\omega_j,{M_0}_j)_{j=1,..,n}$ is
\begin{eqnarray}
    \RV(t) &=& c_0 + c_1 t + \sum\limits_{j=1}^{n} f(t;K_j,P_j,e_j,\omega_j,{M_0}_j).  \label{eq:nominal_model1}  
\end{eqnarray}
The RV amplitude of planet $j$, $K_j$ is proportional to $m \sin i$ (see Eq.~\eqref{eq:semiamp}). \ch{In the absence of information on $i$, the planetary mass $m$ cannot be determined unambiguously. Constraints on $i$ can be obtained} if the planet transits or, in the case of systems with multiple high-mass short-period planets, due to detecting planet-planet interactions \citep[e.g.][]{laughlin2001, correia2010, nelson2016, rosenthal2019}.

When the data are acquired with different instruments, it is important to account for the fact that they might have different zero velocity references. Hence, if there are $m$ different instruments, $c_0$ in Eq.~\eqref{eq:nominal_model1} should be replaced by $\sum_{j=1}^{m} c_0^j \chi_{j}(t)$, where $\chi_{j}(t) =1$ if the measurement at $t$ is taken by instrument $j$ and 0 otherwise.  



\subsection{Basic model of observed data}
\label{sec:sampling}

To detect the variations due to planets, astronomers measure the RV of a given star at several irregularly spaced epochs, usually from 20 to 1,000. 
Each star is only observable when it is high above the horizon, typically for only a few hours per night, and 
weather can prevent stars from being observed.    
Furthermore, a star is typically observable for $\sim$ 4-8 months, as it must not be too close to the Sun.
Finally, most observatories support multiple science programs, 
so observations on a given star might be interrupted for a period from hours to weeks.



\begin{figure}[h]
\includegraphics[width=10cm]{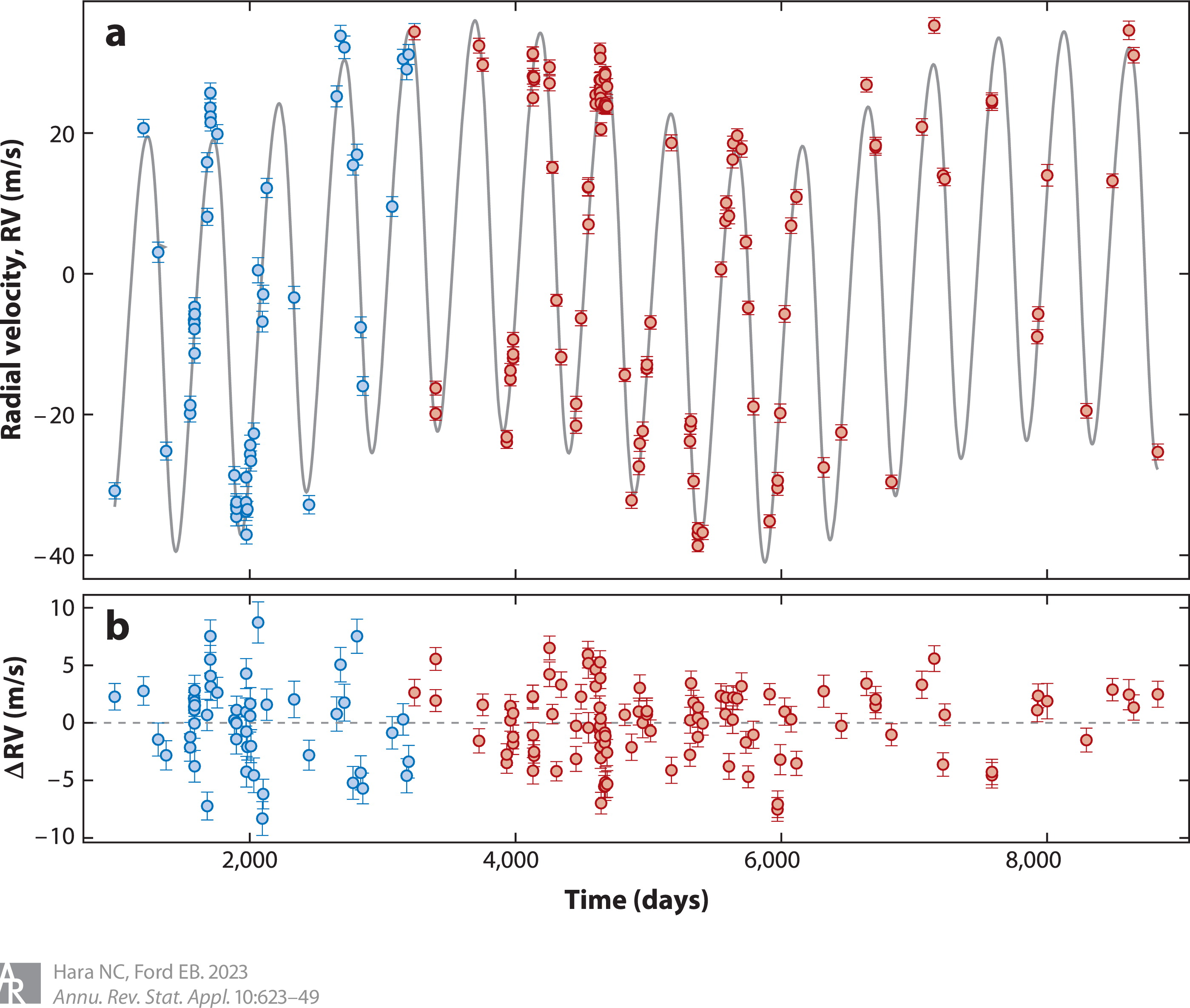}
\caption{(a) Radial velocity (RV) measurements of the star HD 114783. The gray curve shows the maximum
likelihood 2-planet model. (b) RV residuals relative to the 2-planet model above. Error bars reflect reported
measurement uncertainties ($\sigma_{t_i}$) and do not reflect the 3 m/s of jitter ($\sigma_{J}$). Points are color-coded to indicate
whether they are before or after an instrument upgrade that resulted in an unknown RV offset. Data are
from \citet{rosenthal2021}.}
\label{fig:rv_HD114783}
\end{figure}

Each RV measurement has a nominal uncertainty depending on the number of photon received, as well as the properties of star, spectrograph and data reduction technique \citep{bouchy2001}. 
Astronomers typically assume independent Gaussian measurement noise at each time $t_i$, but
in order to reduce the risk of underestimating uncertainties (e.g., unmodeled planets, instrumental noise or stellar variability), an extra Gaussian noise term, often called ``jitter'', is added to the model \citep{ford2006}, leading to 
\begin{eqnarray}
    y(t) & = & \RV(t) + \epsilon(t), \label{eq:generic_model} \\
     \epsilon(t_i) &\sim & N(0,\sigma_{t_i}^2 + \sigma_J^2),  \label{eq:noise_model_indep}
\end{eqnarray}
where $\RV(t)$ is defined in Eq.~\eqref{eq:nominal_model1} and the value of the assumed noise for the measurement made at $t_i$, $\epsilon(t_i)$, follows a Gaussian distribution of variance $\sigma_{t_i}^2 + \sigma_J^2$. 
Fig.~\ref{fig:rv_HD114783} shows an example of a two-planet fit on the RV data of HD114783.





\subsection{Other Processes Affecting RVs}
\label{sec:noise_sources_stellar}

The model in Equations~\eqref{eq:nominal_model1}-\eqref{eq:noise_model_indep} has been extensively used to estimate the masses and orbital parameters of exoplanets \citep[e.g.,][]{wrighthoward2009, bonomo2017b}. 
For favorable stars, it is useful for analysing RV signals with amplitude over $ 3-10$ m/s =  33 - 110 $\mathrm{RV}_\oplus$, enough to characterise short-period planets with masses greater than Neptune.
However, detecting less massive planets requires very precise observations and/or many observations. 
Accurately interpreting their RV signatures requires greater sophistication because multiple potential noise sources become relevant and require more detailed models for $\epsilon_t$. These sources are listed below, and presented in more detail in Appendix \ref{app:understanding_the_signal}.

  \subsubsection{Stellar effects} \label{sec:granulation}
The principle of Doppler spectroscopy is to measure the velocity of a light source with respect to an observer, so if the hot gas in different parts of the stellar surface has different brightness and velocities, it has different impacts on the data. Several physical processes on the surface of stars have an RV signature.

  Convection near the surface of the star creates outward and inward motion of the gas. \ch{This creates a pattern of evolving convection cells. This process, referred to as granulation, has an effect on} RV measurements that can be described as a stationary noise \ch{with a Lorentzian or super-Lorentzian power spectrum} \citep{harvey1985, dumusque2011i, cegla2019, guo2022}. 
  \ch{Granulation effects} on the RV have been simulated in detail \citep{meunier2015, cegla2019, dravins2021}.


 

  \ch{Local enhancement of the magnetic field at the surfaces of stars might result in regions darker or brighter than their surroundings, called spots and faculae, respectively~\citep{schrijver2002, strassmeier2009}.}
  This creates an imbalance in flux from the the approaching and receding side of the \ch{rotating} star, and reduces upward convection, which has an additional net RV effect.  
 The effect of magnetic activity on RV has been studied in \citet{saar1997, dumusque2012ii} and \cite{haywood2016}, and simulated in \citet{lagrange2010a, boisse2012, dumusque2014} and \cite{gilbertson2020}. Active regions grow in size rapidly and disappear more gradually. The visible stellar surface is similar but not identical to itself after one stellar rotation, which results in quasi-periodic RV variations. 
  
  
  
   The rate of appearance, size and location of spots and faculae varies over a $\sim$ 10 year time-scale on solar-type stars, although it can be down to $\sim$ 1 year for M-type stars \citep{corteszuleta2023}.  This affects the \ch{amplitude} of short-term RV variations due to magnetic activity (up to 25 m s\textsuperscript{-1} \citep{lovis2011b}) and changes the net RV effect over these timescales. Appendix \S\ref{sec:stellar} provides a description of known stellar effects affecting RVs, and we refer the reader to \cite{cegla2019_review},  \cite{meunier2021}, and  \cite{crass2021} for reviews.

\subsubsection{Other effects} \ebf{Residual instrumental systematics and absorption of certain wavelengths by the Earth atmosphere can result in spurious RV signatures  of the order of a few $\mathrm{RV}_\oplus$} \citep[e.g.][]{cunha2014, artigau2014, bertaux2014, smette2015}.  \ebf{As they} are not amenable to a concise statistical description, we refer the reader to Appendix \S\ref{sec:calib} and \cite{halverson2016, cretignier2021} for details.

\section{STATISTICAL FRAMEWORK}
\label{sec:indicators}

\subsection{Model of RV}

The planets only affect the Doppler shift, which must be estimated and converted to RV measurements.
We can express the RV \ch{extracted from the spectra}, $\widehat{\RV(t)}$, 
as the sum of 
$\RV(t)$, the RV at time $t$ due to a motion of the center of the mass of the star (in particular due to planets), 
$\RV_\mathrm{contam}(t)$, the RV caused by  stellar variability and instrument systematics, and  
$\epsilon(t)$, the inevitable \ch{photon} noise. Then, we have
\begin{align}
   \widehat{\RV(t_i)} &=  \RV(t_i) + \RV_\mathrm{contam}(t_i)   + \epsilon(t_i) . 
   \label{eq:rv_contribution_contam}
   \\
   \label{eq:rv_contribution}
   \RV(t_i) &= \RV_\mathrm{planets}(t_i) + \RV_\mathrm{g}(t_i), i=1,...,N,
\end{align}
where $\RV_\mathrm{planets}(t_i)$ and  $\RV_\mathrm{g}(t_i)$ are respectively induced by planets and by other gravitational effects such as companion stars or proper motion in the galaxy. \ch{In Eq.~\eqref{eq:nominal_model1}, $\RV_g(t) = c_0 + c_1 t$.} 

\begin{figure}[h]
\includegraphics[width=0.98\linewidth]{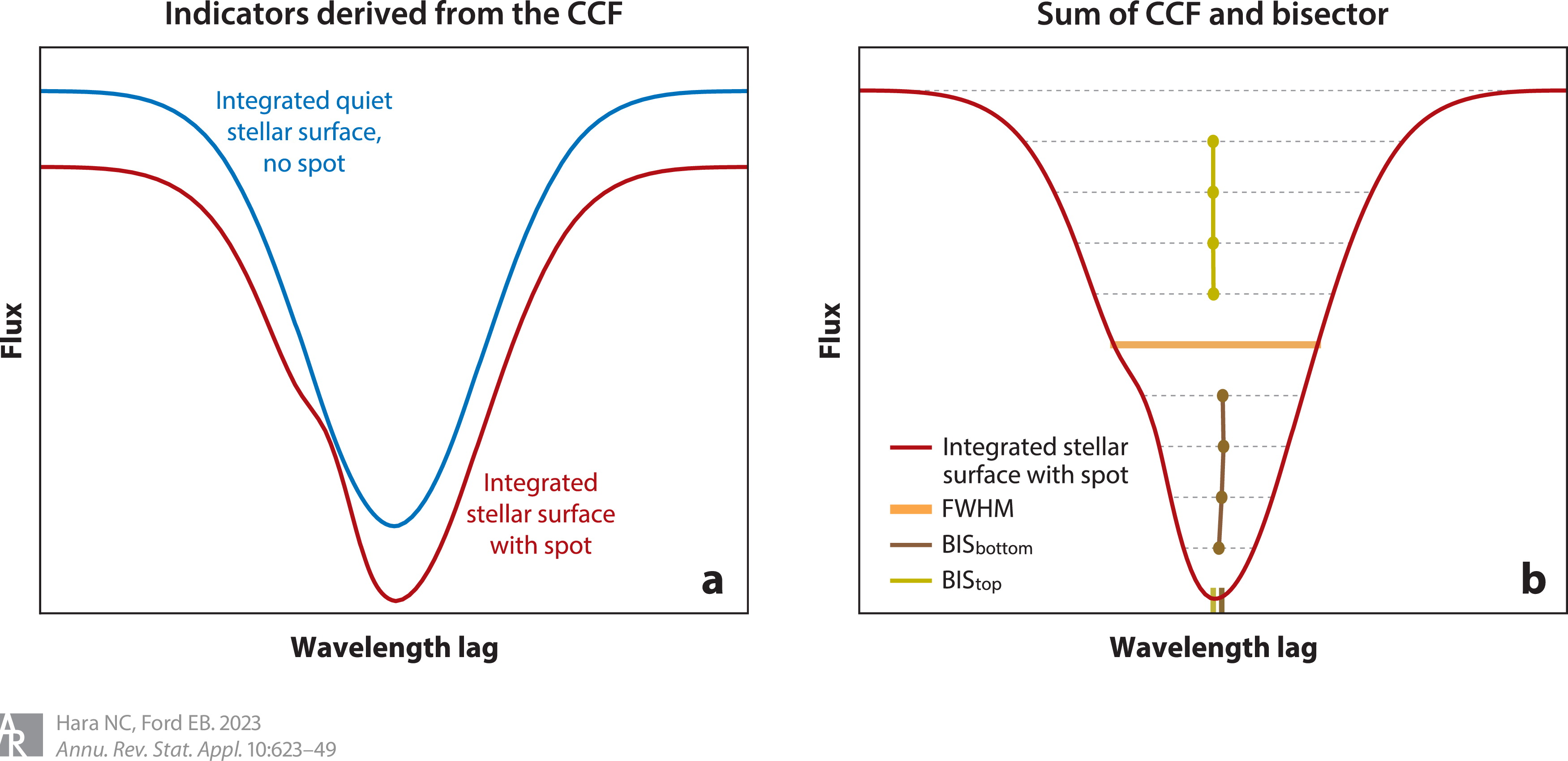}
\caption{Schematic cross-correlation function (CCF). The scalar product of a mask and the spectrum as a function of the wavelength offset between the two, which can be visualized as an average spectral line. (a) The blue curve shows the CCF of the quiet stellar surface, and
the red curve shows the CCF when a dark spot is present on the surface. (b) The CCF when the spot is present has a different full width at half maximum (FWHM) than the CCF of the quiet stellar surface and is asymmetric. The difference between the average position of the lower and upper parts of the CCF, known as bisector inverse slope (BIS) \citep{queloz2001}, is nonzero. As the star rotates, the spot CCF will have a different Doppler shift: The bump in the red CCF moves from left to right, which also causes the shape of the spectra to change with time, in particular the BIS and FWHM. The spot in this figure covers 20\% of the surface, whereas spots on the sun
cover $\sim$0.01–0.1\% of the surface.}
\label{fig:ccf}
\end{figure}

In order to interpret RV measurements properly, we must adequately model $\RV_\mathrm{contam}(t)$. Unlike planetary RV signals, RV variations induced by the star and instrument do not have a constant frequency, phase, and amplitude (see \S\ref{sec:qualitative}), and stellar processes and systematics not only cause a Doppler shift in the spectrum but also cause the shape of the spectrum to vary with time (see Figure \ref{fig:ccf}) and do not affect all lines identically \citep[e. g.][]{thompson2017, wise2018, cretignier2020_lbl, ning2019, wise2022, almoulla2022a, lafarga2023}. The shape variations of the spectra can be used to predict the associated $\RV_\mathrm{contam}(t)$ signal, either in a machine learning approach with a training set of spectra, or in a statistical framework. In the following section, we show how these ideas can be expressed in a general statistical formalism.

%



\begin{marginnote}
\entry{CCF Mask}{Mask: function
of the wavelength
equal to unity except
on wavelength ranges
corresponding to
spectral lines \citep[see][]{pepe2002} }
\entry{Ancillary indicator}{ time series of $N$ scalars extracted from time series of $N$ spectra, summarizing a shape variation of the spectrum, for instance the variations of the average spectral line, or of the amplitude of a specific line.}
\end{marginnote}

\subsection{Challenges of RV data analysis}
\label{sec:challenges}

Our understanding of exoplanets can be thought of as a language where each planet is a word, and each planetary system is a sentence. As in information theory~\citep{shannon1948}, our knowledge of this language comes through a noisy and biased communication channel. In the long term, we want to understand the full language, but first we must learn to interpret each sentence: what planets orbit a given star? The Bayesian formalism is well suited to describe such situations, and indeed it offers a compact way to present the different problems of RV data analysis \citep[][also uses an information-theoretic framework for exoplanets]{sandford2019}.

\ch{In our analogy, the message received is the raw data.} The most fundamental data product is the number of electrons counted on each CCD detector of the spectrograph.\footnote{We could go even further and consider data from an entire survey of spectral time series of different stars as a population using a hierarchical model to better constrain instrumental effects and potentially stellar variability.} 
\ch{The sentence to be decoded is represented by vectors $\utheta$ and $\ueta$. The vector $\utheta = (\utheta_1,...,\utheta_n$) represents the planetary system, where $n$ is the number of planets (which can also be treated as an unknown variable) and $\utheta_j$ the orbital elements of planet indexed by $j$ (see Eq.~\eqref{eq:rv_planet}). 
The vector $\ueta$ includes all other relevant non-planetary parameters, such as an offset, a drift, the stellar rotation period, noise amplitude and timescale.} 



\ch{Let us denote the lowest-level data by $\vec D$. Our goal is to obtain a meaningful joint posterior distribution for $\ueta$ and $\utheta$ knowing $\vec D$, $p(\utheta, \ueta \mid \vec D)$, but $\vec D$ is corrupted by hundreds of instrumental effects, and computing the posterior directly is currently unmanageable. To be usable $\vec D$ needs to be reduced to an estimate of the RV time series $\vec{\widehat{\RV}} = (\widehat{\RV}(t_i))_{i=1,..,N}$ and ancillary indicators $\mat I = (\vec I_j(t_i))_{j=1..p, i=1..N}$, such that for each $j$, $\vec I_j(\vec t)$ is a  time series of length $N$, summarizing the variations of shape of the spectrum (see Fig. \ref{fig:ccf}.b). } With notations of Eqs. \eqref{eq:rv_contribution_contam} and \eqref{eq:rv_contribution},
%
\begin{align}
    p(\utheta, \ueta \mid \vec D) \approx p(\utheta, \ueta \mid  \mat I, \widehat{\mathbf{RV}})  &= \frac{p( \mat I, \widehat{\mathbf{RV}} \mid \utheta, \ueta) p( \utheta, \ueta)}{p(\mat I, \widehat{\mathbf{RV}})} = \frac{p( \mat I, \mathbf{RV}_{\mathrm{contam}} + \uepsilon \mid \utheta, \ueta) p( \utheta, \ueta)}{p(\mat I, \widehat{\mathbf{RV}})}.
    \label{eq:approx_posterior}
\end{align}
The last equality comes from the fact that $\RV(t)$ depends deterministically on $\utheta, \ueta$. Below, we denote $\vec y = (\widehat{\mathbf{RV}}, \mat I)$, \ci{potentially with $\mat I = \emptyset$}. We will refer to $p(\utheta, \ueta)$ as the prior distribution and $p(\vec y \mid \utheta, \ueta)$ as the likelihood. In some cases, to emphasize that the likelihood is restricted to models with exactly $n$ planets we will denote the likelihood by $p(\vec y \mid \utheta, \ueta, n)$.  This formalism also accounts for the case where photometric data of the target star is also available: the photometry time series can be considered as another indicator, with the added complexity that the data is not acquired at the same epochs as the spectra.
\begin{marginnote}
\ch{ \entry{Reduced data}{vector
y concatenating the
RV time series and
ancillary indicators,
extracted from the raw
data $\mathbf{D}$}

\entry{Likelihood}{probability
distribution of the
reduced data $\vec y$
knowing the model
parameters $\utheta $ and $\ueta $;
denoted by $p(\vec y \mid \utheta, \ueta)$}
}
\end{marginnote}


Based on Eq.~\ref{eq:approx_posterior}, we can identify several choices to be made: 
$(a)$ a \textit{reduction} method transforming the observational raw data into summary statistics $\widehat{\mathbf{RV}}, \mat{I}$ that estimate the RV and provide useful indicators of \ch{spectral} variability; 
$(b)$ a \textit{model}, that is a likelihood $p(\vec y \mid \utheta, \ueta)$ and prior distribution $p(\utheta, \ueta)$. The key difference with step $(a)$ is that the effect of potential planets is now explicitly included in the likelihood definition;  
$(c)$ a \textit{decision}  method for confidently claiming the detection of planetary signals and estimating their masses and orbital elements for a given model, or a collection of models. \ch{In the Bayesian formalism above, the decision is based on the posterior distribution $p( \utheta, \ueta \mid\vec y)$} and must include a discussion of its sensitivity of the result to the choices made in $a$ and $b$. The Bayesian formalism only serves here to compactly present points $a$--$c$ and is not always used in practice.
Points $a$, $b$, and $c$ are treated recursively in \S \ref{sec:analysis_methods_dl}, \ref{sec:GP_activity} and  \ref{sec:analysis_methods_ts}, respectively, and must be completed by $(d)$ a set of practical and reliable \textit{numerical methods} for performing the necessary calculations. Numerical aspects are highlighted when relevant.




\section{DETECTING PLANETS AND ESTIMATING THEIR ORBITAL ELEMENTS}
\label{sec:analysis_methods_ts}



%



\subsection{Likelihood}
\label{sec:likelihood}

Suppose that we have at our disposal an RV time series, a few ancillary indicators and a statistical model of them. We want to determine how many planets can be confidently detected, and what their orbital elements are.


\ch{In \S\ref{sec:indicators}, we presented the different steps of exoplanet detection, and in particular of a likelihood function describing the distribution of data $\vec y$ as a function of the model parameters $\utheta$ and $\ueta$, describing respectively the planets and all other effects.} The likelihood is commonly assumed to be Gaussian,
 \begin{equation} 
    \mathcal{L} \equiv p(\vec y \mid \utheta, \ueta ) = \frac{ \e^{-\frac{1}{2}  \left[\vec y - \vec  g(\vec t; \utheta, \ueta) \right]^T \mat {V(\ueta)}^{-1} \left[\vec y - \vec g(\vec t; \utheta, \ueta) \right]}}{\sqrt{(2\pi)^N  |\mat V(\ueta)|} } .
    \label{eq:likelihood_simple}
\end{equation}
where $|\mat V(\ueta)|$ is the determinant of the covariance matrix $\mat V(\ueta)$. 
For instance, supposing that $\vec y$ is the RV time series, using model~\eqref{eq:nominal_model1}-\eqref{eq:noise_model_indep},  $ \utheta =  (K_j,P_j,e_j,\omega_j,{M_0}_j)_{j=1,..,{n}}$, $\vec g$ is the sum of Keplerians and an affine function (see Eq.~\eqref{eq:nominal_model1}), and $\ueta = \sigma_J, c_0, c_1$, $\mat V$ is diagonal with $i$-th element $\sigma_{t_i}^2 + \sigma_J^2$. 
If $\RV_{\mathrm{contam}}$ is non-zero, it might be modelled by a non-diagonal matrix $\mat V(\ueta)$, and/or a linear model of stellar activity indicators in $\vec g$. This form of the likelihood is used both when the data $\vec y$ is the RV time series only, a single ancillary indicator, or the concatenation of the time series of RV and ancillary indicators (possibly including also photometric observations of the target star).
We return to how to specify further $\vec g(\vec t; \utheta)$ and $\mat V(\ueta)$ in \S\ref{sec:GP_activity}.




Assessing the statistical significance of a putative RV signal 
is remarkably challenging. We review three broad approaches: periodogram-based methods (\S\ref{sec:model_comparison_periodogram}), Bayesian model comparison (\S\ref{sec:model_comparison_bayesian}), and in \S\ref{sec:qualitative}, we present approaches that do not aim to specify an explicit model for nuisance signals, but aim to obtain robust detections.

\subsection{Planet Detection via Periodograms}
\label{sec:periodsearch}

The decision on the planets detected and their orbital elements should make use of the full posterior of orbital elements, or in a frequentist setting the full likelihood. However, 
computing the posterior distribution of elements given in Eq.~\eqref{eq:approx_posterior} \ch{or exploring exhaustively the parameter space to evaluate the likelihood} is only possible with the latest numerical tools, and is computationally intensive. Realistic datasets often have \ci{millions of} local likelihood maxima. Historically, the exploration of likelihood modes was done with periodogram methods, which are still extensively used due to their speed, numerical stability, and ability to unveil the dominating frequencies in the data.
\subsubsection{Definition}

Given a time series, for instance the RV time series, or the time series of a given ancillary indicator, periodograms consist in comparing the log-likelihoods
of two models: a base model $H_0$ and a model $K_\omega$ including $H_0$ plus a periodic component at fre-
quency $\omega$, for a grid of frequencies. In periodograms, the Keplerian model is often replaced with
an approximation that yields a convex model once conditioned on a small number of parameters.

The first and simplest case is the  Lomb--Scargle periodogram \citep{lomb1976,scargle}, where the base model $H_0$ is Gaussian white noise, and $K_\omega$ is the same white noise plus a sine function. Both $H_0$ and $K_\omega$ are described with likelihoods such as  that~\eqref{eq:likelihood_simple}, with 
\begin{align}
H_0: \; &\vec g  = \vec{0}, \label{eq:h0}\\ 
K_\omega :\; & \vec g(A,B,\omega)  = A\cos \omega  \label{eq:komega} \vec t + B\sin \omega \vec t, 
\end{align}
and $\mat V = \mathrm{diag}((\sigma_i^2)_{i=1,...,N})$ in both models.
In $K_\omega$, $\vec g$ is the exact RV signature of a planet on a circular orbit ($e=0$) with orbital period $P = 2\pi/\omega$ and is a good approximation when $eK \leq \sigma_{\mathrm{RV,ideal}}$. For a given $\omega$, maximizing the likelihood with respect to $A$ and $B$ \ch{is equivalent to} to minimizing the sum of squares, and it is a linear problem. The Lomb--Scargle periodogram is then the difference of the log-likelihoods of models $H_0$ and $K_\omega$ as a function of $\omega$.

\begin{figure}
\includegraphics[width=11cm]{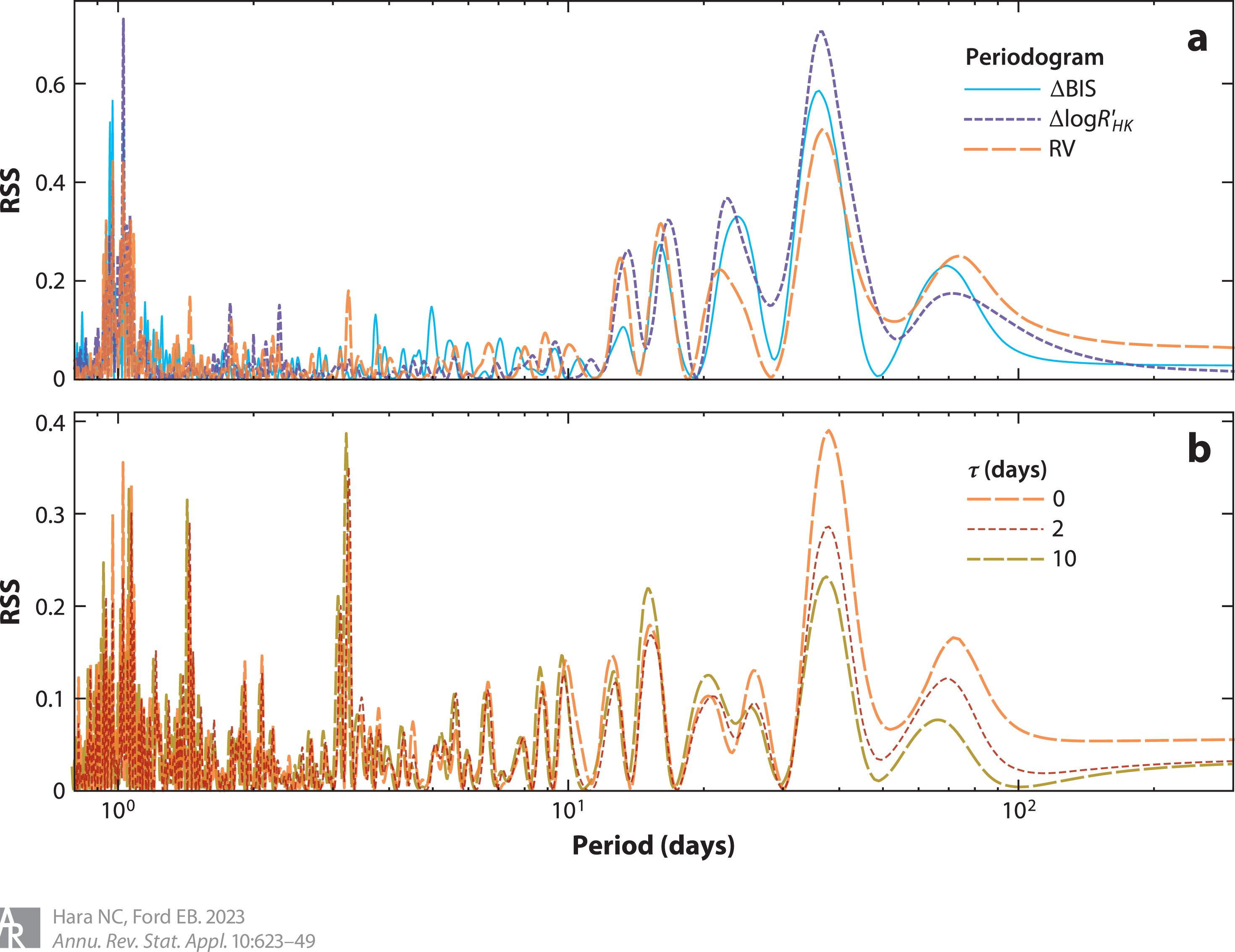}
\caption{(a) generalized Lomb--Scargle periodograms of the time series of two ancillary indicators: bisector inverse slope,  $\Delta$BIS \citep{queloz2001}, and $\Delta\log R'_{HK}$ \citep{noyes1984}), and RV time series (RV) of the ESPRESSO (Echelle SPectrograph for Rocky Exoplanets
and Stable Spectroscopic Observations) data of the star TOI 178178~\citep{leleu2021}. The time series of $\Delta$BIS, $\log R'_{HK}$ and RV are shown in Fig.~\ref{fig:gps}.  Periodograms are expressed in reduction in sum of squares (RSS), normalized --- that is, the difference between residual sum of squares after fitting $H_0$ and $K_\omega$, divided by that of the $H_0$ residuals (see Eqs.~\eqref{eq:h0} and~\eqref{eq:komega}). The periodogram presents peaks around a period of 1 day, due to the repetition of observations every 24 hours $\pm$ 2-3 h~\citep{dawsonfabricky2010}.   
All three time series present power at $\sim40$ days, due to stellar rotation. (b) In orange, periodogram of the RV time series when the $\Delta\log R'_{HK}$ and $\Delta$BIS are included as a linear \ci{predictors} in the $H_0$ and $K_\omega$ models. This damps the amplitude of the stellar rotation: / signal, and a peak at 3.2 days emerges. The red and light brown curves correspond to adding a correlated noise Gaussian model in the covariance with a Gaussian kernel of amplitude 1 m/s and timescale 2 or 10 days. Adding correlated noise models always damps power at low frequencies (for details, see \cite{delisle2019a}). }
\label{fig:periodograms}
\end{figure}

The principle of the periodogram can be extended with more complex definitions of $H_0$ and $K_\omega$ and/or by interpreting the periodogram in a Bayesian context. The null hypothesis $H_0$ can be complexified \citep[e. g.][]{baluev2008}, the periodic signals might be chosen as nonsinusoidal \citep[e. g.][]{baluev2013_vonmises, baluev2015}, and the assumption that the noise is uncorrelated can be dropped \citep[e.g.][]{delisle2019a}. Figure \ref{fig:periodograms} shows an example of the application of different types of periodograms. A comprehensive list of existing periodograms is given in Appendix \ref{app:periodogram}.


\subsubsection{Periodogram-based model comparison}
\label{sec:model_comparison_periodogram}

The usual way to determine whether a periodogram supports the detection of a periodic signal is to compute $p( \max_{\omega}\mathcal{P}(\omega) \mid H_0)$ the probability distribution of the \ci{maximum value of the periodogram, $\max_{\omega}\mathcal{P}(\omega)$} under a null hypothesis, $H_0$. 
We compute the maximum of the periodogram for the data to be analyzed, $\mathcal{P}_{\mathrm{d}}$ \ci{on a grid of frequencies $\Omega = (\omega_i)_{i=1..N}$},
and define a false alarm probability, FAP, as $p(\max_{\omega \in \Omega}\mathcal{P}(\omega) \geqslant \mathcal{P}_{\mathrm{d}} \mid H_0)$. %

The FAP is most robustly estimated by generating signals under the null hypothesis, and the FAP estimated is the fraction of simulations with a maximum peak greater than $\mathcal{P}_\mathrm{d}$.
This is computationally expensive, as it requires $\sim n$ simulations for a precision of $100/n\%$ on the FAP. 
Since astronomers often aim for a FAP of  $<$ 0.1\% to 0.01\%, $n>10^4$ is needed. 

A semianalytical approach allows one to simulate a reduced number of signals and to fit a generalised extreme value distribution \citep{suveges2014} to the empirical distribution. 
The analytical approach approximates the FAP thanks to the theory of extreme values of stochastic processes using the Rice formula~\citep{baluev2008, baluev2009, baluev2013}, even for complex periodic shapes~\citep{baluev2013_vonmises, baluev2015} or correlated noise~\citep{delisle2019a}. These analytical approximations are very
accurate in practice
if at least a few tens of data points are available \citep{suveges2015}. 


\vspace{-0.3cm}

\subsubsection{Pros and cons}  

Periodograms are fast and numerically stable, and the advent of accurate
analytical estimates of the FAP, even in the case where signals are searched for simultaneously
 \citep{baluev2013} or with Gaussian correlated noise models \citep{delisle2019a}, further simplifies their use. They provide a useful visual diagnostic of the frequency content of the signal. However, they have drawbacks.

First, most periodograms scan for the orbital period of one planet at a time. For multiple planet systems, periodograms can be applied iteratively, first to identify the most conspicuous planet from the data, then to look for a $(j+1)$th planet in the residuals of the best-fit model using $j$ planets.   
However, periodograms are sensitive to aliasing, or spectral leakage, \citep{dawsonfabricky2010}. 
Aliases of different signals 
can add destructively or coherently, so, even if the highest peak of a periodogram is statistically significant, its period may not match that of any physical signal.  
This can be avoided by searching for multiple signals simultaneously \citep{ford2011, baluev2013}. 
Brute force search for two signals at once is expensive and allowing for more rapidly becomes prohibitive. Other approaches based on sparse recovery allow one to search for several signals at low computational cost \citep{hara2017}.

%

Stellar activity causes low-frequency signals which can be mistaken for planets. It is better not
only to search for several planets simultaneously but also to fit the parameters of stellar activity
and planetary signals jointly.
As discussed in \S\ref{sec:likelihood}, stellar variability is often modeled as correlated RV noise using a non-diagonal covariance matrix $\mat V$ that requires additional parameters that must be inferred from the data.  
This means that the log likelihood is non-convex for a given orbital period.  
Marginalizing (or even optimizing) over the kernel parameters dramatically increases the computational cost of periodograms, although it can be done \citep{delisle2018}.

\subsection{Bayesian Approach to Planet Detection}
\label{sec:model_comparison}
\label{sec:model_comparison_bayesian}

A more principled approach to comparing models with different numbers of planets is to directly use the Bayesian formalism of Eq.~\ref{eq:approx_posterior}. The first method consists in computing the the Bayesian evidence, also called marginal likelihood, of the $n$-planet model. Letting $\Theta_n$ denote the parameter space of all possible combinations of $n$ planets,
\begin{equation}
    p(\vec y \mid n) = \iint_{\Theta_n} p(\vec y \mid \utheta, \ueta, n ) p(\utheta, \ueta\mid n) \dd \utheta \dd \ueta . 
    \label{eq:evidence}
\end{equation}
where $p(\vec y \mid \utheta, \ueta, n )$ is given by Eq.~\ref{eq:likelihood_simple}. 
If one can compute the Bayesian evidences for models with $n+1$ and $n$ planets, then their ratio gives the ``Bayes factor'' \citep{kassraftery1995}.
If the noise model is correct, then when one considers more planets than are justified for the given data set, the evidence decreases \ch{as more planets are added}. \ch{For instance, when the priors on the different parameters are considered independent, the prior term decreases geometrically with the number of planets} but the model does not result in a significantly higher likelihood. 
Bayesian model comparison for exoplanet detection was suggested by~\cite{gregory2005a, gregory2007a, fordgregory2007, tuomi2009} and \cite{ford2011}, and has become one of the primary methods for establishing the statistical significance of \ch{detections}. 

 \ch{Astronomers compute the Bayes factors for increasing $n$, starting at $n=0$, until they are below a certain threshold.  }
The literature contains various heuristics for the interpretation of Bayes factors \citep[e.g.,][]{jeffreys1961}.
Unfortunately, following guidelines blindly can be very dangerous.
For important scientific discoveries astrophysicists routinely demand that a frequentist test rejects a null hypothesis test with a $p$-value of $\sim~10^{-3}$ or even $\sim~10^{-7}$ before publishing a result.  
\ebf{In such situations, a Bayesian would demand a posterior odds ratio (i.e., prior odds ratio times the Bayes factor) exceeding $10^3$ or even $10^7$ before publishing a discovery.}   \ch{A Bayes factor strongly favoring an $(n+1)$-planet model over an $n$ planet model, does not necessarily imply that there must be an $(n+1)$th planet, if the physical or statistical models used are inaccurate. It is important to check the dependence of the results on the adopted model.}



%
Bayes factors compare different numbers of planets. It is insufficient to claim a confident detection of a planet, which requires knowledge of the orbital elements to a certain accuracy: there is little value in a planet detection without estimates of its size and period. Well-defined orbital
elements correspond to a sharp posterior mode, whose existence can be checked with the posterior
samples of orbital elements. This can be done with periodograms, or directly with the posterior
distribution of orbital elements if it can be estimated reliably. If so, the detection criterion can be constructed to convey information about the planets' location. For instance a detection claim can be defined as: ``there is a planet with orbital elements in $\Theta$'', where $\Theta$ is a region of the parameter space \citep{brewer2015, handley2015,hara2021a}. In that case, the criterion minimizing the number of missed detections for a certain tolerance to false ones is the false inclusion probability~\citep[FIP,][]{hara2022}, i. e. the probability of not having a planet with orbital elements in $\Theta$.

A significant barrier to more routine adoption of Bayesian model comparison is the difficulty of computing the Bayesian evidence accurately. \cite{nelson2020} compared different methods to compute the Bayesian evidence and found that the agreement between them goes from a factor $\sim 1$ for 0 planet models to $\sim 10^2$ for 2 planet models. Additionally, the observed dispersion of Bayes factor estimates from multiple runs of the same method was often significantly greater than the reported uncertainties for some methods.  There, they recommend evaluating the numerical uncertainty based on several independent runs.  When a fast estimate is needed, a Laplace approximation is advised over heuristics such as AIC or BIC.  Numerical details are discussed further in Appendix \ref{sec:computations}.   


\subsection{Qualitative approach to planet detection} 
\label{sec:qualitative}

The methods presented in \S\ref{sec:periodsearch} and~\ref{sec:model_comparison_bayesian} rely on a complete model of the signal. All the alternative hypotheses are made explicit and are compared to one another. This gives meaningful statistical significance indicators and measures of uncertainty. However, if none of the noise models captures stochastic variation in the data, the inferences are unreliable.  A second approach  consists, in the formalism of Eq.~\eqref{eq:rv_contribution}, of extracting meaningful information about $\RV_\mathrm{planets}(t)$ without explicitly specifying a model for $\RV_\mathrm{contam}(t)$. Such approaches also might serve another purpose: diagnosing unanticipated effects. 


\ch{On the timescale of RV observations, unless there are strong gravitational interactions between the planets, planetary signals are purely periodic, unlike for stellar and instrumental signals. To diagnose whether a signal is truly periodic} \cite{schuster1898} and \cite{mortier2017} compute classical periodograms by adding one point at a time and checking that the amplitude of the peak corresponding to the candidate planet increases steadily. Alternatively, \cite{gregory2016, hara2021b} use the Bayesian framework described in \S\ref{sec:model_comparison_bayesian} and adds \ebf{an} \ebf{apodization factor} to the Keplerians: the model of  Eq.~\ref{eq:rv_planet} is multiplied by a Gaussian term $\e^{-(t-t_0)^2/(2\tau^2)}$, where $t_0$ and $\tau$  are free parameters. If the signal is consistent, the probability that  $\tau$ exceeds the total observation time-span should be high. 

Another line of work consists in searching for periodic signals in the data, without specifying a parametric form. \cite{zucker2015} suggests using a Hoeffding test. \cite{zucker2018} applies the formalism of distance correlation to evaluate the statistical dependence of a cyclic variable with period $P$ and the data, and this is further explored by \cite{binnenfeld2021}, who aim to find RV variations which are statistically independent from the spectral shape variations. This work uses spectra information through  all pairwise distances between two spectra measurements, and shows promising results.

\vspace{-0.3cm}

\subsection{Parameter estimation and uncertainty quantification}
\label{sec:param_est}

Once the dominant local modes of the likelihood or posterior have been identified by a periodogram analysis or with a random sampler with good convergence properties, more computationally intensive methods can be employed for finer parameter estimation in this region. The posterior distribution of the elements can be evaluated with MCMC algorithms \citep{ford2005,gregory2005}.
Brute-force random walk MCMC requires carefully chosen proposal distributions for systems with up to four planets \citep{ford2006}. 
Modern studies typically use ensemble samplers \citep{nelson2014, foremanmackey2013}, adaptive Metropolis sampling~\citep{delisle2018}, Hamiltonian and/or geometric MCMC samplers \citep[e.g.,][]{papamarkou2021}, or pre-marginalisation with a Laplace approximation of the evidence over the linear parameters~\citep{pricewhelan2017}.
In each of these methods, the orbital periods are initialised at multiple values very near the dominant signals found by the periodogram analysis. 
The results are generally reliable if the likelihood is dominated by a single mode and the initial estimate of the period falls into that mode. 
Testing numerical convergence when using MCMC or other iterative algorithms.  Some nested samplers have shown good performances in blindly locating different likelihood maxima \citep{brewer2015, handley2015b, handley2015, faria2022}.

\section{MODEL: SPECIFYING THE PRIORS AND LIKELIHOODS}
\label{sec:GP_activity}

Different choices of priors and likelihoods have a very strong impact on the detection of planets. Priors have a strong influence both on the detection and orbital elements estimation of low amplitude signals, where the likelihood is less constraining \citep{hara2021a}. To avoid priors that
are unrealistically diffuse, a possibility is to use well tested reference priors and to exclude orbital
configurations that are unstable \citep{tamayo2020, stalport2022}. A discussion of the influence of priors and a list of commonly used ones is provided in Appendix \ref{sec:bayesian_priors}, and we now focus on the likelihood, $p(\vec{RV} \mid \mat I, \vec D)$ (see Eq. \eqref{eq:approx_posterior}). 



Historically, the first method was to represent RV variations induced by the instrument and star as a linear combination of the ancillary indicators, $\mathbf{RV}_{\mathrm{contam}} = \sum_j \alpha_j \vec I_j$ \citep{queloz2001, queloz2009, dumusque2016}. This neglects the fact that indicators are themselves noisy. The efficiency of this method is also limited due to the fact that there \ebf{can be} phase shifts between activity-induced changes in RVs and ancillary indicators, reducing the correlation~\citep{santerne2015, lanza2018, colliercameron2019}.  For rotational-linked variability, it can be useful to generalize to models that predict $\RV_{\mathrm{contam}}$ at time $t$ using  spectra taken at times near $t$ \citep{colliercameron2019,zhao2022}.

\begin{figure}[h]
\hspace{-2cm}
\includegraphics[width=13cm]{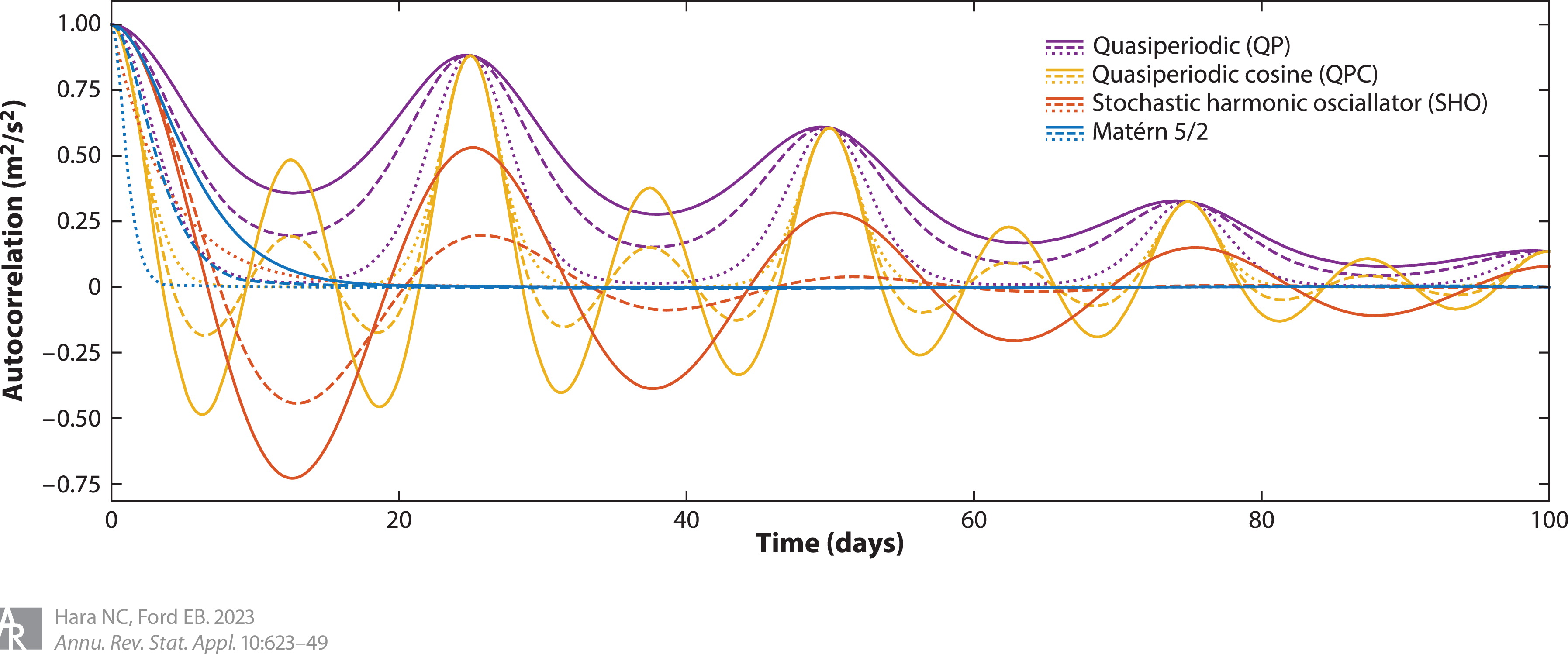}

\vspace{0.3cm}
\caption{Kernel functions: functions of the time lag $\Delta t$ between two observations. The covariance matrix $\mat V$ in Eq.~\eqref{eq:likelihood_simple} is such that its element $i,j$ is a sum of kernel functions evaluated in $\Delta t = t_i - t_j$. Each color corresponds to a different family of kernels: Quasi periodic~\citep[QP][]{aigrain2012, haywood2014}, quasi-periodic cosine~\citep[QPC][]{perger2021}, stochastic harmonic oscillator~\citep{foremanmackey2017}, Matérn 5/2~\citep{foremanmackey2017, gilbertson2020}. Each family has several parameters, including an amplitude, a time-scale, and for all but Matérn 5/2, a period. They have additionally shape parameters and plain, dashed and dotted correspond to different values of this shape parameters. The periods are taken as 25 days, close to the Sun rotational period, and all kernels are normalized so that they are equal to 1 in $\Delta t = 0$. These different kernels model different phenomena (see \S\ref{sec:granulation}). The QP kernel is well  suited to represent stellar rotational effects due to convective blueshift inhibition with various stellar inclinations. QPCs model the fact that spots appear preferentially on stellar longitudes with a 180$^\circ$ phase shift in Sun-like stars~\citep[QPC][]{borgniet2015, spergel2015}. SHO kernels have take negative values, this type of anticorrelation arises due to the breaking of imbalance of flux of the approaching and receding limb, and certain parameter values correspond to a super-Lorentzian power spectral density used to model granulation~\citep{foremanmackey2017, gilbertson2020}. Matérn 5/2 are efficient tio model RV induced by spots and faculae with a lifetime lower than the rotation period~\citep{gilbertson2020}. Explicit kernels expression and values of the parameters are given in Appendix \ref{sec:stellar_variability_as_correlated_noise}. }
\label{fig:kernels}
\end{figure}

Another way to account for the signals corrupting RVs is to represent them as correlated Gaussian noise. One can still follow the formalism of Eq. \eqref{eq:likelihood_simple} assuming that here $\vec y $ is the RV time series, \ebf{but now specifying} the matrix $\mat V$ with a kernel, which gives the correlation between the value of the stellar RV signal at $t$ and $t+\Delta t$. A list of common kernels is provided in Appendix \ref{sec:stellar_variability_as_correlated_noise}. The correlation of RVs due to stellar variability can be expressed as a sum of the correlations due 
to each of the processes described in Section 2.4 (see Fig. \ref{fig:kernels}). However, using a Gaussian process representation of the RV only has serious limitations: it acts as a frequency filter, and thus might spuriously decrease the significance of viable planet candidates. To use 
the information contained in the ancillary indicators, account for their random nature and complex relationships to one another, recent work use Eq. \eqref{eq:likelihood_simple} but where the data $\vec y$ is the concatenation of RV and ancillary indicators time series in the framework of Gaussian processes.


Gaussian processes were introduced to the exoplanet community in \cite{aigrain2012}. They showed that \ebf{when nearly contemporaneous} photometric and RV observations of a star are available, the effect of one stellar spot on RV can be predicted through the photometric flux and its derivative. For most targets, continuous photometric measurements are
unavailable. One can estimate the derivative of the flux at any time using Gaussian processes
(GPs). These are stochastic processes $G(\vec t)$, function of a variable $\vec t$ such that for any $n$ values of $\vec t$, $\vec t_1,...,\vec t_n$, $(G(\vec t_1),...,G(\vec t_n))$ follows a multivariate Gaussian distribution. In general, $\vec t$ can be a vector, but for our purpose, we define it as the time $t$. This implies that the GP is defined by two quantities: its mean $m(t)$ and kernel, $k(t,t')$, equal to the covariance of $G(t)$ and $G(t')$ \citep{rasmussen2005, aigrain2022}.



\begin{figure}[h]
\includegraphics[width=14cm]{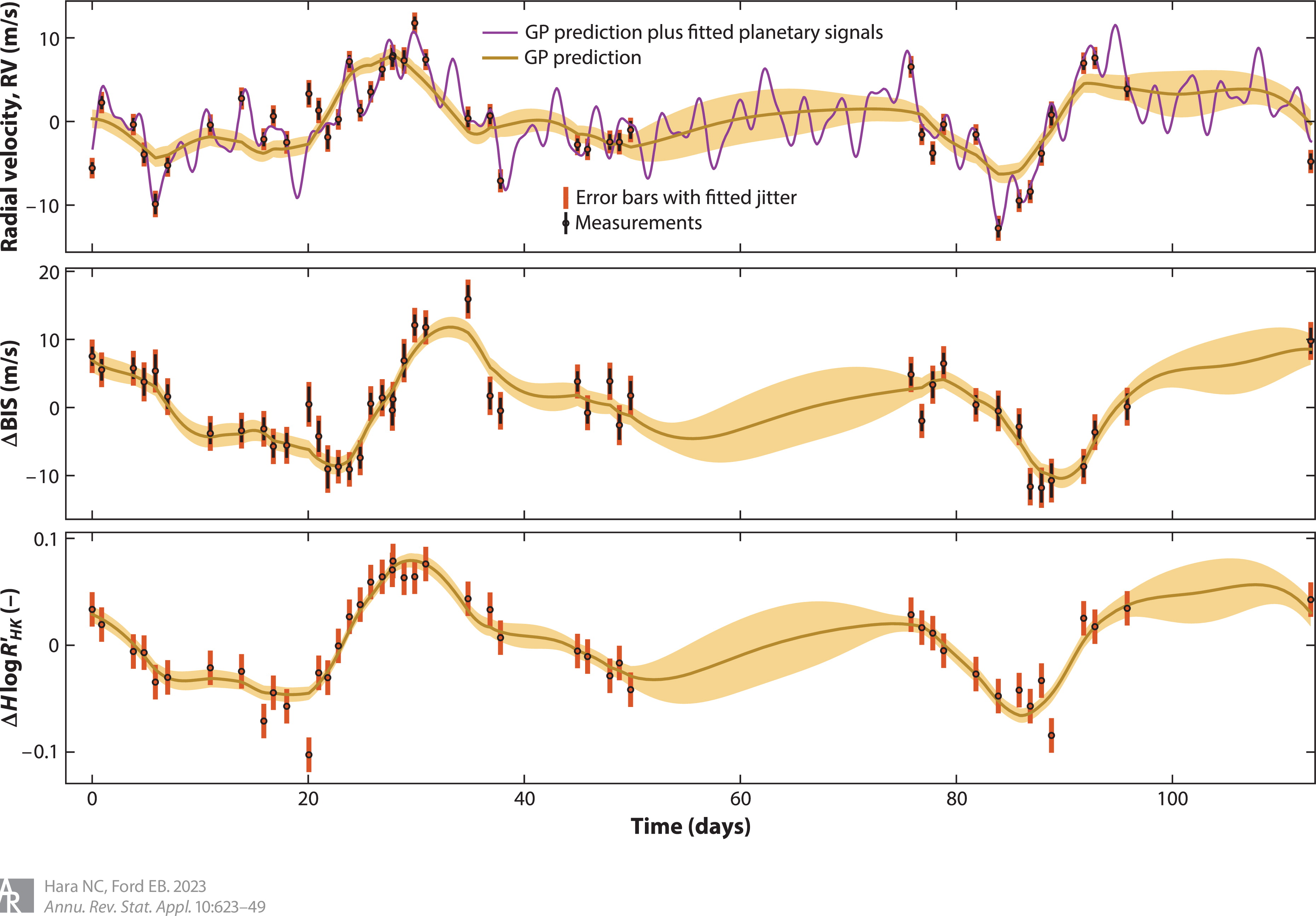}
\caption{\ch{Gaussian process regression applied to the time series of RV, and ancillary indicators $\Delta$BIS~\citep{queloz2001} and $\Delta\log R'_{\mathrm{HK}}$~\citep{noyes1984} of 46 measurements obtained on the star TOI 178 with the ESPRESSO instrument~\citep{leleu2021}. 
Each of the three time series $y_i(t)$, $i=1,2,3$ is represented by a model $y_i(t) = \alpha_i G(t) + \beta_i \dot{G}(t) + \epsilon_m(t) + \epsilon_J(t)$. The Gaussian process $G$ has a stochastic harmonic oscillator (SHO) kernel~\citep{foremanmackey2017} with three parameters, fitted with the S+LEAF package~\citep{delisle2022}. The processes $\epsilon^m(t)$ and $\epsilon^J(t)$ represent the measurement error and jitter. We represent with a brown line the mean of the posterior predictive distribution for each time series, and the shaded yellow areas represent $\pm$1 standard deviation of $\alpha_i G(t) + \beta_i \dot{G}(t)$. The purple curve represents the variations predicted
by the best-fit Keplerian model for the planet, whose presence is confirmed by transits \citep{leleu2021}. Other abbreviations: BIS,
bisector inverse slope; $\Delta\log R'_{\mathrm{HK}}$, ancillary indicator derived from each spectrum.
}
}
\label{fig:gps}
\end{figure}

\cite{aigrain2012}  use photometric data to predict the RV variation. We can go a step further and analyze RV and ancillary indicators simultaneously by building a likelihood $p(\vec{RV},\mat I \mid \vec D)$, \ebf{typically} in the Gaussian process (GP) framework.  RVs and ancillary indicators are expressed as linear combinations of one or several \ebf{latent} Gaussian processes and their derivatives \citep{rajpaul2015, gilbertson2020, gordon2020, barragan2022, delisle2022, tran2023}. Since $G(t+\Delta t) \approx G(t) + \dot{G}(t) \Delta t $, using the derivative \ebf{can} account for small phase shifts between RV and indicators. The data $\vec y$ in Eq.~\ref{eq:likelihood_simple} is then the concatenation of the RV vector and ancillary indicators.  Fig.~\ref{fig:gps} shows an example of RV and ancillary indicators modeled simultaneously with a GP and its derivative. 

Representing RV and ancillary indicators as a linear combination of a GP and its derivative is only one way to specify the covariance in Eq.~\ref{eq:likelihood_simple}, other options have been proposed. For example, the GP framework can be used to model simultaneously the RVs derived in different spectral bands~\citep{cale2021}. The kernel of the extended data $\vec y$ can also be based on explicit physical models of the star \citep{luger2021_II, luger2021_III, hara2023}. Furthermore, the assumption that the likelihood is Gaussian and stationary can be relaxed. \cite{hara2023} quantifies non Gaussianity with an analytical model of stellar variability. \cite{camacho2023} model the RV and ancillary indicator time series in a GP regression network. It allows, in particular, consideration of noise with heavy tails and offers ways to account for non-stationarity due to the magnetic cycle. 


Evaluating the likelihood Eqn.~\ref{eq:likelihood_simple} for certain values of the parameters requires the log determinant of $\mat V(\ueta)$ and computing $\{\vec y - \vec  g(\vec t; \utheta)\}^T \mat V^{-1}(\ueta) \{\vec y - \vec g(\vec t; \utheta)\} $.  With standard algorithms the computational cost scales as $O(N^3)$ with the size $N$ of the dataset $\vec y$, and becomes prohibitive for $N\gtrsim 10^3$. This prevents using many indicators on stars with hundreds of points, or analysing the tens of thousands of RV measurements on the Sun.
Fortunately, for certain choices of the kernel, the covariance matrix has a \ebf{``semi-separable''} expression and then the computational cost scales as $O(N)$ \citep[the CELERITE framework,][]{foremanmackey2017, foremanmackey2018}.  Similarly, \texttt{TemporalGPs.jl} \footnote{\url{https://github.com/JuliaGaussianProcesses/TemporalGPs.jl}} allows for $O(N)$ computation of a broad class of 1-d kernel functions in Julia.  
The CELERITE framework is generalised by~\cite{delisle2019b} to the S+LEAF framework, which models covariances as a sum of semi-separable and \ebf{``LEAF matrices''}\footnote{\url{https://gitlab.unige.ch/Jean-Baptiste.Delisle/spleaf}}. Quasi-separable kernels provide even greater flexibility for modeling multivariable timeseries (\texttt{tinygp} \footnote{\url{https://tinygp.readthedocs.io/en/stable/index.html}}). 
\cite{delisle2022} show that if RV and ancillary indicators are linear combinations of a Gaussian process $G(t)$ and its derivatives, and the covariance of $G(t)$ has a S+LEAF form, the computation of the likelihood of the augmented data is still $O(N)$\footnote{ \url{https://gitlab.unige.ch/Jean-Baptiste.Delisle/spleaf}}.


The merits of the different noise models can be evaluated through Bayesian model comparison (see \S\ref{sec:model_comparison_bayesian}), as in \cite{ahrer2021, faria2022, suarezmascareno2022}, alternatively Bayesian model averaging can be used \citep{hara2022}.




\section{ANALYSIS METHODS: DEEPER LEVELS}
\label{sec:analysis_methods_dl}

So far, we have seen different methods to detect planets and estimate their orbital elements given priors and likelihood in $\S$\ref{sec:analysis_methods_ts}. We have seen how the model can be built based on a certain reduction of the data in $\S$\ref{sec:GP_activity}. We now present  techniques to extract \ch{physical} information from the spectra. 

Our goal is to identify conceptual similarities between different methods, rather than to assess their performances. 
While the community has begun to compare methods~\citep{zhao_expres_2022}, \ch{proper evaluations are} challenging since one does not know the true RV for real stars due to potential undetected planets \citep[except for our Sun;][]{colliercameron2019, dumusque2020, lin2022}. We suggest metrics to compare approaches in \S\ref{sec:conclusion}.

\subsection{Estimating RV}

For a fixed stellar spectrum, the RV is well defined and can be extracted from the spectrum using multiple techniques described below. However, at the sub-m/s (\ci{or a few $\mathrm{RV}_\oplus$}), level, the shape of the spectrum changes, so, there is no unambiguous definition of RV, and each method to estimate it relies on more or less explicit assumptions.
Depending on how this shape change is modelled, the estimate of the Doppler shift changes. 


Three main families of methods for extracting the velocity in each spectrum separately are in use. As discussed in \S\ref{sec:indicators}, the historic method consists of cross-correlating the spectrum with an idealised mask \citep{baranne1979, pepe2002}. Denoting by $f(\lambda)$ the flux as a function of the wavelength $\Lambda$,
\begin{align}
    \mathrm{CCF}(\Delta \lambda) = f \ast \mathrm{Mask} (\Delta \lambda)
\end{align}
where $\ast$ is the convolution operator. 
This method can be viewed as computing an average line shape.
Each line contributes to improving the SNR, but also contributes a bias since true line wavelengths are unknown.
Additionally, the CCF loses information about differences in shapes of each line.  
Often, many lines are excluded from the CCF mask to reduce contamination of lines likely to contribute significant bias. Alternatively, \cite{lienhard2022} proposes a least-square deconvolution technique to estimate a common spectral line profile. 

A second approach, template matching, consists in measuring the RV based on a template or model spectrum and a Taylor expansion for the spectrum as a function of velocity \citep{connes1985, bouchy2001, angladaescude2012, astudillodefru2015, jones2022, silva2022}. Related work has been done by \cite{cretignier2022}. This approach is based on the approximation
\begin{align}
    f( \lambda)  \simeq f_0(\lambda) + \frac{RV}{c} \frac{\dd f_0}{\dd\log\lambda}\left(\lambda\right),
\end{align}
where $f_0(\lambda)$ is a reference spectrum 
and $c$ is the speed of light.
Both the CCF and template matching approaches  aim to reduce the impact of stellar variability by careful selection of lines/wavelengths for inclusion, but lack a mechanism to recognize stellar variability. Some authors measure the RV of each line separately~\citep{dumusque2018, artigau2022} (or $\sim~2$\AA ``chunk'' of the spectrum), and compute a weighted average RV after a statistical rejection of outliers, or do independent RV estimates in several bandwidths \citep{feng2017, zechmeister2018}. The rationale is to disentangle signals which affect all wavelengths identically (planet candidates) from other signals.

A third approach builds a forward model of the spectrum, 
\begin{align}
    f( \lambda^o) =    \left[ f_0(\lambda^{bc}) + \frac{RV}{c} \frac{df_o}{d\log\lambda}( \lambda^{bc} ) \right] T(\lambda^o) \ast IP(\lambda^o),
\end{align}
where\ci{$f(\lambda^0)$ is the spectrum} in the observer frame, $\lambda^{bc}(t) = \lambda^o \{1+v_{bc}(t)/c\}/\sqrt{1-v_{bc}^2(t)/c^2}$ is the wavelength in a frame that accounts for $v_{bc}(t)$, the known motion of the observatory relative to the solar system barycenter (the ``barycentric correction''), 
where $T(\cdot)$ is an atmospheric transmission profile, and $IP(\cdot)$ is the instrument response.
Variations on forward modeling have \ch{focused on telluric effects
\citep{butler1996,hirano2020}} \ch{or on stellar variability \citep{bedell2019,gilbertson2020,jones2022}}.  Ongoing research is developing computational tractable approaches for including both simultaneously \ch{(e.g., StellarSpectraObservationFitting.jl)}.  

As methods become more sophisticated, the estimation of the velocity and fitting of spectral shape tend to be made simultaneously on the time series of spectra. In principle, this mitigates the chance that the Doppler shift is contaminated by shape changes, and uses temporal structure to further constrain activity.



\subsection{Estimating nuisance RV signals }
\label{sec:nuisance sig}

To put Eq.~\eqref{eq:approx_posterior} into practice, one must specify a statistical model for $p(\mat I, \mathbf{RV}_{\mathrm{contam}} \mid \utheta, \ueta)$.  First, one chooses the method of dimension reduction to obtain $\vec I$, and then a form for the likelihood which describes $\mat I$ and $\mathbf{RV}_{\mathrm{contam}}$. 



Dimension reduction has been applied to real and synthetic spectra.
\citet{davis2017} applied PCA to spectra generated with SOAP 2.0.  Since real stellar spectra are more complicated, their analysis provides a lower limit on the number of PCA components needed to accurately reconstruct solar spectra (ranging from 1 to 4 depending on the spectral resolution and SNR).  
Analysis of solar data suggests that only 6-13 basis vectors are necessary to model solar variability at the resolution and SNR of HARPS-N observations, and at least four of those are clearly linked to effects due to the instrument or unique to sun-as-a-star observations \citep{colliercameron2021}.  Together, these suggest that reducing spectra to an RV and two to six indicators is a fruitful direction for future research.

Dimension reduction and estimation of contaminating RV can be done simultaneously.  Existing methods extract ancillary indicators either in each spectrum separately, from a time series of CCFs (or other summaries of the spectra), or from the time series of spectra themselves. 
In the following subsections we describe the associated methods.




\subsubsection{Extraction spectrum by spectrum}

Some methods extract activity indicators from each spectrum separately. For example, one can measure line shapes (or deviations from their time-average).  
Shape line indicators were first extracted from the CCF, in particular its asymmetry \citep{queloz2001} and width \citep{queloz2009}. Other indicators involve fitting Gaussian distributions on each side of the CCF \citep{figueira2013}. \cite{holzer2021} fit a superposition of Gauss--Hermite functions to each line to build shape indicators. 

Another approach is to measure properties of specific absorption lines of interest.
The utility of different lines depends on the effective temperature of the host star.  
For sun-like stars, $\log R'_{HK}$ and the so-called $S$-index are popular magnetic activity indicators based on emission in the core of the Calcium II H\&K lines \citep{noyes1984, wright2004}.  
For cooler stars, the $H\alpha$ line is a more useful indicator \citep{kurster2003, robertson2016}. 
The unsigned magnetic field estimated from the spectrum can be a powerful indicator of $\RV_{\mathrm{contam}}$ \citep{haywood2020} and might generalize better than any individual line.

\cite{santerne2015} and \cite{lanza2018} find \ci{that activity indicators can have relatively weak correlations with RV signals contaminating the data.}  
One contributing factor is the expected time lag between RVs and other indicators, noted in the case of RV and photometry \citep{queloz2001, santos2003, queloz2009} and other indicators \citep{suarezmascareno2017}. This time lag can be partially mitigated in models where the derivative of the latent Gaussian process appears~\citep{rajpaul2015, delisle2022} or with adaptive correlation \citep{simola2022}.


\subsubsection{Methods using stellar variability indicators}

More generally, spectra can be reduced to a set of summary statistics that serve as stellar variability indicators and their time series can be analyzed jointly with the RVs.  By utilizing temporal information, we may obtain less noisy indicators.

The most common reduction is to compute the CCF.
However, good indicators should be insensitive to true Doppler shifts,     
so
\citet{colliercameron2021} proposed \ch{{\sc Scalpels}} to analyze the autocorrelation of the CCF. The resulting time series of CCF autocorrelations are analyzed with a PCA and the PCA scores are used as variability indicators. 
Since the autocorrelation function is insensitive to shifts, true Doppler shifts will not affect the resulting variability indicators.  
\cite{debeurs2020} explored more flexible supervised learning approaches to predicting $\RV_{\mathrm{contam}}$ from a training data set, either simulated or solar data CCFs.  
It is unclear whether the greater flexibility of neural networks will outweight their added complexity and difficulty of interpretation relative to linear regression or {\sc Scalpels}, particularly given the limited size of datasets available for stars other than the Sun.  

\cite{cretignier2022} transform the spectra into ``shells'' instead of CCFs, in an effort to reduce the amount of information lost when averaging lines of different depths. 
The PCA scores for the shell are used as spectral indicators, but are first orthonormalised with respect to the shell corresponding to a pure Doppler shift. 

More research is needed to develop effective means of comparing choices of summary statistics, variability indicators and corresponding likelihoods.  Several methods were explored in the EXPRES Stellar Signals Project, where several teams analysed observations of four stars by EXPRES \citep{zhao_expres_2022}. One key finding was that while many methods could reduce the \ci{root mean square} RMS RV of observations, the estimates of $\RV_{\mathrm{contam}}$ differ significantly across methods.  
Without knowing the true velocity, it was impossible to determine which methods are best. \ci{In \S\ref{sec:future_work}, we present several ideas for a robust comparison of the different methods leveraging the existing and upcoming RV observations of the Sun with different instruments.}

Sun-as-a-star observations \ebf{allow the testing and validation of methods} \ebf{for mitigating stellar variability.}
\ebf{Three methods, {\sc Scalpels} \citep{colliercameron2021}, linear regression on CCFs and neural networks \citep{debeurs2020} have been shown to significantly reduce the level of RV variability in solar observations.}
\ci{Existing and} upcoming comparisons of Sun-as-a-star observations from multiple instruments, including a new generation of more highly stabilized spectrographs, will help disentangling solar RV variations due to the solar variability and from instrument specific signals, and refine the comparisons of different analysis methods.  


\subsubsection{Methods using the time series of spectra}

\citet{rajpaul2020} propose an alternative approach of measuring a $\Delta \RV$ from each pair of spectra and reconstructing the RV time series (modulo a constant), while giving greater weight to pairs that are more similar.  
In order to reduce the computational cost, they split the spectra into many small chunks to be analyzed separately. 

In principle, one could perform inference on the entire spectroscopic time series.  
\citet{jones2022} introduced Doppler-constrained PCA on the time series of spectra to characterize stellar variability and provided a proof of concept on simulated solar observations.  
\cite{bedell2019} applied a similar method to real observations, emphasizing modeling of telluric contamination and neglecting stellar variability.  They regularized the variables using $\ell_1$ and $\ell_2$ norm constraints.
More recent work has developed 
\ch{computationally efficient} implementations that simultaneously model both stellar and telluric variability, as well as accounting for the instrumental profile and allowing for physically-informed regularization schemes.


\section{CONCLUSION}
\label{sec:conclusion}

\subsection{Summary}

RV is poised to play a key role in the study of exoplanets in the next decade as a primary way
to measure their masses; to detect interesting, nontransiting planets; to have a more complete
view of planetary system architectures; and to provide targets whose atmospheres will be further
characterized by spectro-imaging or interferometry.  Currently, characterizing an Earth twin is out of reach, and improving data analysis techniques will play a fundamental role.

We presented the different steps of RV data analysis, separating them in three problems (see
Section 3): ($a$) reducing the information of the spectrum into an RV time series, ($b$) modeling
nuisance signals and the prior information on planetary and nuisance parameters, and ($c$) deciding
how many planets are present and what their orbital elements are. Each of these steps requires
numerical methods, where convergence should be carefully checked. Since there are multiple
reasonable choices for $a$, $b$, and $c$, researchers should perform multiple analyses with different
assumptions to determine whether key conclusions are sensitive to these choices, especially of
the likelihood and prior. 






\subsection{Future work}
\label{sec:future_work}

The exoplanet community is now well-equipped to analyze RV observations when planetary signals dominate stellar \ci{and instrumental} variability. 
When these planetary signals and corrupting ones are of similar amplitude, the fact that different methods yield different results \citep{zhao_expres_2022} shows that more work remains to be done. We believe that further research in steps $a$ and $b$ is crucial. 

The ability of RVs to detect and measure the mass of potential Earth twins is critically linked to
how precisely the contaminating RV signals, the RV signal not due to the motion of the center of
mass of the star ($\RV_{\mathrm{contam}}$ in Equations \eqref{eq:rv_contribution_contam}-\eqref{eq:rv_contribution}), can be predicted and how accurately the uncertainty
on these predictions can be quantified. Correcting these signals can be done either in a statistical
framework by building a likelihood function or with supervised learning techniques trained to
predict the RV contamination signal from spectral shapes. The new models may be data-driven or
physics-driven, especially for RV signals originating from the star, for which there is a substantial
modeling effort (see Appendix \ref{app:stellar}) yet to be translated to data analysis techniques. Further research in step a is also required to ensure that all the information contained in the spectra is used and
that the RV and indicators derived are reliable summary statistics.


 
\ci{We believe that one of the most critical question} is how to validate choices for \ch{each} step 
as effective tools for detecting and characterizing low mass planets in the presence of stellar variability.  We propose three ways forward, the first two concerning the validation of reduction and modelling (step $a$ and $b$), and the third the whole process ($a$--$c$).

The first approach is Bayesian model comparison, to \ch{evaluate the relative merits of} different stellar activity and instrument systematics models. Ideally this would be applied to datasets with no ambiguity on the presence or not of planetary signals.  
For example, since we know the true solar velocity, we can evaluate the signal $\mathrm{RV}_{\mathrm{contam}}$ of each method applied to sun-as-a-star observations. A similar approach may be possible for some other stars hosting a massive planet on an eccentric orbit, since one can exclude the possibility of additional planets for a wide range of orbital periods based on orbital stability considerations \citep{brewer2020, stalport2022}. 
Large and computationally demanding computer models could provide data for training and testing models, though they must be validated beforehand. \ci{The observations of the Sun by different high precision spectrographs can be leveraged to disentangle instrument specific noise from stellar variability.}

Second, for stars where one acquires a sufficiently large number of observations, one could evaluate the accuracy and precision of a model mitigating stellar variability via its ability to predict $\RV_{\mathrm{contam}}$ (see Eq.~\eqref{eq:rv_contribution_contam}) and stellar variability indicators at times not used in the training of the model.  Given multiple processes operating on a variety of timescales, one must carefully design the training, testing and validation procedure.  
For example, if one obtained multiple spectra per night and stellar variability were operating on timescales of weeks, then one could trivially predict the $\RV_{\mathrm{contam}}$ from another observation on the same night, without learning how to recognize stellar variability in line shapes or depth ratios. Additionally, one must be careful that cross validation is not undermined by researchers  effectively trying many strategies and reporting results from those that appear to work best. 

The two approaches above pertain to validating models for contaminating signals. The most convincing method for validating models may be via planet injection-recovery tests.  
\ch{One group of researchers would work on injecting planets in real or simulated datasets.}
Another group of researchers would blindly analyze large ensembles of simulated datasets. 
 This would allow each proposed data reduction method and likelihood to be evaluated \ch{differentially}, i.e., comparing the inferred velocities from multiple simulated data sets generated from the same true data set. 
Given the inevitable arbitrary nature of labeling some signals as ``confident'' exoplanet detections, it is not sufficient for teams to label which signals they believe is due to an exoplanet.  Instead, we recommend that they provide a list of all putative signals along with a quantiative measure of the signals' statistical significances (FAP, Bayes factor, FIP, see \S\ref{sec:analysis_methods_ts}). Then, the number of false detections can be computed as a function of the number or properties of missed planets. To generate the data, one approach would be to remove telluric and instrumental effects, introduce a larger number of artificial Doppler shifts, reinject the telluric and instrumental effects, and generate many new synthetic data sets (perhaps adding additional noise along the way). One must be careful in implementing this approach, as an error in the injection process could become a feature learned by data-driven methods that would not be available for realistic data sets.

In this review, we adopted the viewpoint that RV data analysis should be viewed as a whole from the most basic data products to the final decisions. It will be crucial to build accurate instrument systematics and stellar noise models, to leverage as much as possible the information contained
in the spectra, and to build reliable metrics to validate the different methods. Progress will most likely stem from the combination of a deep knowledge of the instruments and astronomical context combined with formal approaches, which will provide adequate tools to represent and analyze the
data.
 

 
 \section*{ACKNOWLEDGMENTS}
The authors thank Michaël Crétignier \ch{who suggested presenting the} RV reduction methods in three broad classes and his comments, Sahar Shahaf, Xavier Dumusque, Suzanne Aigrain and Charles Cadieux for their insightful suggestions, \ci{as well as David Hogg for suggesting to compare the prediction errors of different methods}. N. C. H. acknowledges the financial support of the National Centre for Competence in Research PlanetS of the Swiss National Science Foundation (SNSF). \ch{E. B. F. acknowledges the financial support  from the Heising-Simons Foundation Grant \#2019-1177.
The Center for Exoplanets and Habitable Worlds is supported by the Pennsylvania State University and the Eberly College of Science.}

\newpage

\appendix

\Large

\textbf{SUPPLEMENTAL APPENDIX}

\normalsize

\section{DEFINITIONS}
\label{sec:definitions}

For convenience we reproduce here the formalism used in the \textbf{main text}. We denote by $\RV(t)$ the radial velocity (RV) of a given star due to a motion of its center of the mass at time $t$. We denote by
$\RV_\mathrm{contam}(t)$ the RV signal caused by stellar variability and instrument systematics, and  
$\epsilon(t)$, the \ch{photon} noise. From an estimated Doppler shift of the spectrum, the radial velocity measurement estimated at time $t$, $\widehat{\RV(t)}$, is
\begin{align}
   \widehat{\RV(t)} &=  \RV(t) + \RV_\mathrm{contam}(t)   + \epsilon(t) . 
   \label{eq:rv_contribution_contam_app}
   \\
   \label{eq:rv_contribution_app}
   \RV(t) &= \RV_\mathrm{planets}(t) + \RV_\mathrm{g}(t)
\end{align}
where $\RV_\mathrm{planets}(t)$ is the RV signal caused by the planets and $\RV_\mathrm{g}(t)$ is the RV signal caused by other gravitational effects, such as the proper motion of the star in the galaxy, or the influence of another star. The time series of radial measured velocities is denoted by $\widehat{\mathbf{RV}} = (\widehat{\RV}(t_i))_{i=1,..,N}$.

Let us suppose that a star is observed by a spectrograph at $N$ different times, $t_i, i=1,..,N$. We denote by $\vec D(t_i)$ the lowest level data product at time $t_i$, the fluxes measured on the spectrograph detector of the  and various housekeeping data (temperature, pressure etc). In principle, we could base our inference on the time series $\vec D = (\vec D(t_i))_{i=1,..N}$. Loosely speaking, we want to use non Doppler variations of $\vec D$ to estimate as closely as possible $\RV_\mathrm{contam}(t)$, in particular the variation of the shape of spectral lines, and color-dependent effects. We can also leverage the fact that, unless there are strong dynamical interactions between planets, the planetary signals are purely periodic, but systematics and stellar activity are not. 

We denote by $\utheta$ the parameters of the planets orbiting the star and $\ueta$ all other parameters describing the signal (offset, noise level, parameters of the noise kernel), respectively.  We have $\utheta = (\utheta_1,...,\utheta_n$) where $n$ is the number of planets (which is also a variable) and $\utheta_j$ the orbital elements of the planet indexed by $j$ (see Eq.~\eqref{eq:rv_planet_app}). $\vec D$ is reduced to an estimate of the RV time series $\widehat{\mathbf{RV}}$ and ancillary indicators $\mat I = (\vec I_j(t_i))_{j=1..p, i=1..N}$, are $p$ time series of length $N$, summarizing the variations of shape of the spectrum. The goal is not to loose too much information in the reduction process. We thus want
\begin{align}
    p(\utheta, \ueta \mid \vec D) \approx p(\utheta, \ueta \mid  \mat I, \widehat{\mathbf{RV}}), 
    \label{eq:approx_posterior_app}
\end{align}
and thanks to the Bayes formula, we have
\begin{align}
p(\utheta, \ueta \mid  \mat I, \widehat{\mathbf{RV}}) = \frac{p( \mat I, \widehat{\mathbf{RV}} \mid \utheta, \ueta) p( \utheta, \ueta)}{p(\mat I, \widehat{\mathbf{RV}})} = \frac{p( \mat I, \mathbf{RV}_{\mathrm{contam}} + \uepsilon \mid \utheta, \ueta) p( \utheta, \ueta)}{p(\mat I, \widehat{\mathbf{RV}})}.
\end{align}
The last equality comes from the fact that $\RV(t)$ depends deterministically on $\utheta, \ueta$. Below, we denote $\vec y = (\widehat{\mathbf{RV}}, \mat I)$, \ci{potentially with $\mat I$ = 0}. We will refer to $p(\utheta, \ueta)$ as the prior distribution and $p(\vec y \mid \utheta, \ueta)$ as the likelihood. The notation $p(\vec y \mid \utheta, \ueta, n)$ refers to the posterior restricted to models with exactly $n$ planets. 

The likelihood is commonly assumed to be Gaussian,
 \begin{equation} 
    \mathcal{L} \equiv p(\vec y \mid \utheta, \ueta ) = \frac{ \e^{-\frac{1}{2}  \left[\vec y - \vec  g(\vec t; \utheta, \ueta) \right]^T \mat {V(\ueta)}^{-1} \left[\vec y - \vec g(\vec t; \utheta, \ueta) \right]}}{\sqrt{(2\pi)^N  |\mat V(\ueta)|} } 
    \label{eq:likelihood_simple_app}
\end{equation}
where $|\mat V(\ueta)|$ is the determinant of the covariance matrix $\mat V(\ueta)$. If $\vec y = \widehat{\mathbf{RV}}$ (and we assume non-interacting Keplerian orbits), then the vector $\vec g = (g(t_i))_{i=1..N}$ is such that 
\begin{align}
    g(t, \ueta, \utheta) = \sum\limits_{j=1}^{n} f(t;K_j,P_j,e_j,\omega_j,{M_0}_j) + h(t;\ueta).  \label{eq:nominal_model1_app}  
\end{align}
for some user-defined function $h(\ueta)$, for instance intercepts (also called offsets) modelling the uncertainty on the reference 0 velocity of the instruments. The function $f$ is defined as  
\begin{eqnarray}
f(t;K,P,e,\omega,M_0)  & = & K\left[\cos\left(\omega + \nu(t;e,P,M_0)\right) + e \cos \omega\right] \label{eq:rv_planet_app}
\end{eqnarray}
where
\begin{align}
\cos \nu & = \frac{\cos E - e}{1 - e\cos E} \label{eq:cosnu2bis}\\
\sin \nu & = \frac{\sqrt{1-e^2}\sin E}{1 - e\cos E} \label{eq:sinnu2bis}\\
E - e \sin E &= M_0 + \frac{2 \pi}{P} t \label{eq:keplereq2bis}.
\end{align}
We treat the case where $\vec y$ is a concatenation of RV and ancillary indicator time series in~\S\ref{app:multidim_gp} of this supplemental appendix.

\section{UNDERSTANDING THE SIGNAL}
\label{app:understanding_the_signal}

In \S\ref{sec:calib}, we present briefly how the RV and activity indicators are extracted from the raw data.  In the formalism of \S\ref{sec:definitions}, we present how the data is treated to go from the raw measurements $\vec D$ to 1-dimensional spectra, and from there to activity indicators and RV $\vec y= (\vec I,  \mathbf{RV})$. In \S\ref{sec:stellar} we present the efforts undertaken to understand stellar signals. For exoplanet detection, this can be seen as building knowledge to express a realistic likelihood $p(\vec y \mid \utheta, \ueta)$. Practical likelihoods are given in \S\ref{sec:stellar_variability_as_correlated_noise}.

\subsection{Early stages of signal processing: calibration, wavelength solution, reduction}
\label{sec:calib}

To measure the radial velocity of a given star, its light is collected by a telescope, and, in modern high-precision instruments, an optical fiber in the focal plane of the telescope is connected to the spectrograph.  (In some earlier instruments, a slit was used instead of fibers, but these are being phased out for high-precision radial velocity surveys.)  
The spectrograph will diffract the input light and record the obtained spectrum on a detector (e. g. CCD arrays for spectrographs operating in the visible). Depending on their position, pixels receive light at a certain wavelength. More precisely, to measure radial velocities, a certain design of spectrograph called échelle spectrographs is used. Such a design has the particularity of dispersing the light both vertically and horizontally, which allows to obtain very high resolution spectra with a wide spectral range on a small detector, therefore reducing costs of the detector array and cryostats to cool down the detector. 
The raw data product of a single measurements are flux measured on a two-dimensional CCD arrays.  The extraction of the radial velocity and activity indicators follows many steps. The objective of the first steps is to compute a one-dimensional spectrum: the flux of the star as a function of wavelength. From one exposure to the next, the response of the instrument will vary because of the variations of the environmental conditions of the instrument (temperature, pressure), which varies even for the most recent instruments put in vaccuum and controlled in temperature. Accounting for these changes relies on so-called calibration steps, where the response of the instrument is evaluated with reference lamps which properties are known. There are two types of lamps: one emitting white light (a tungsten lamp, a laser-driven light source, LDLS, or continuum laser), and one with reference emission spectral lines at known wavelength, such as a ThAr or Uranium lamps, Fabry-Perot interferometer or a laser frequency comb. On a Fabry-Perot interferometer, we do not know the position of the lines but the line spacing is known, and their drift on a timescale of a few hours is very low, so we can use such a lamp to measure drift over the course of one night.

Presenting in detail the different steps of the calibration process is well beyond the scope of the present work, and we refer the reader to \cite{hatzes2016} for an introduction (see also \cite{baranne1996, sosnowska2015, fischer2016, brahm2017, petersburg2020, cook2022}). We here briefly describe important steps, which might be performed differently from one instrument to the next: (1) Removing the values of the ``bad pixels'' with aberrant values; (2) Correcting for the ``flat field'': The detector is illuminated with a uniform white light to locate the pixels which receive flux, and correct for the different response of pixels). (3) Extracting the orders: not all the pixels of the detector receive photons from the star. This step includes extracting the position of the pixels receiving these photons and modeling any cross contamination due to light from another order, fiber or scattering in the instrument; (4) Computing the wavelength solution:  The wavelength of the photons falling on each pixel has a certain wavelength that depends on its position. 
 Prior to the observations of the target star, the detector is illuminated with the calibration lamp with reference lines. The pixels corresponding to the emission lines are now attributed a wavelength, and the wavelength of pixels in between are associated to a wavelength via an interpolation with a polynomial or more sophisticated techniques \citep[e.g.][]{dumusque2018, cersullo2019, zhao2021_excalibur, dumusque2021}; (5) Measuring the drift of the instrument: Between the wavelength solution step and the observation of a given star, the wavelength solution of the instrument has changed. To evaluate it, close to the pixels receiving the stellar flux, other pixels are illuminated through a second fiber with a reference lamp. This is done both prior to the observations taken on a given night when the wavelength solution is computed, and during the exposure to the star of interest, and the differences are used to evaluate the shift of the instrument; and (6) Barycentric correction:  The Earth rotates and moves in the Solar system, resulting in a Doppler signature with an amplitude of $\simeq30$ km/s. This contribution is estimated from an ephemeris of the Solar system to cm/s accuracy and accounted for by transforming the wavelength solution to refer to the wavelength in the solar system barycenter frame (neglecting gravitational redshift from the sun) \citep{kanodia2018, tronsgaard2019, blackman2019}.   See \cite{halverson2016} for a review of the error budget corresponding to the steps outlined above.

Once the spectrum is extracted, the spectral energy distribution is affected by the black-body profile of the star, and different observing conditions (airmass, clouds, optical filter ageing, reduced fiber transmission and efficiency CCD response in the blue portion of the spectrum). A step of so-called "color correction" is performed to scale the observed continuum toward a reference template  \citep[see e.g.][]{malavolta2017} A similar color correction can be obtained by normalizing the stellar continuum  \citep[see e.g.][]{xu2019, cretignier2020} --- a step which is unnecessary when extracting the velocity line by line \citep{dumusque2018, artigau2022} ---  before re-injecting a reference color. From the one dimensional spectrum, the radial velocity is extracted with different methods, described in section 6 of the \textbf{main text}. The one dimensional spectrum is still affected by many instrument systematics, the absorption of the Earth atmosphere at certain wavelength, and sunlight scattered by the moon \citep[e.g.][]{bonomo2010, roy2020} which have an important effect on the data. There is much to gain by better understanding these effects. Once an effect has been suspected to occur, one can compute the velocities before and after the procedure aimed at correcting of the said effect, and the difference of velocities gives the contribution of the effect. This methodology has been used in \cite{cunha2014, artigau2014,kjaersgaard2021, allart2022, wang2022} for the correction of the Earth atmosphere absorption, and \cite{cretignier2021} for other various effects. In the latter case, the systematic effects can be seen on a plot in color code of spectra stacked onto one another. 

In \S2 of the \textbf{main text}, we separated the analysis of radial velocity in three steps: (a) Reduction:  From the raw spectrum, we obtain an estimate of the radial velocity and of other stellar variability indicators. These indicators can be  the time series of one dimensional spectra at the available wavelenghts themselves \citep{bedell2019}, or in the majority of cases, a few summary statistics; (b) Modelisation:  We build a likelihood model of the indicators, and possibly attributing them prior distributions;  and (c) Parameter estimation \& Decisions: Based on these information, we decide which planets have been detected and characterize the masses and orbital elements of each detected planet. Typically, it is also important to retriev some information about the host star (often from separate astronomical techniques), since the stellar properties effect the planet mass and climate.  
In the present section, we have seen that step (a) can be in fact subdivided in obtaining the one-dimensional spectrum, then applying further corrections, and finally extracting indicators. 
Similarly, step $c$ combines the detection of planets and the estimation of their orbital elements into one logical step.  In practice, these are often implemented separately for computational convenience.

Overall, the goal of step (a) is two-fold. First, we aim to obtain stellar variability  indicators and RV estimates that have the intended meaning. In particular, the RV that we measure should result from a pure Doppler shift of the spectrum. Second, the information contained in the spectrum should be condensed with as little loss as possible in the RV time series and indicators.

\subsection{Stellar signal}
\label{sec:stellar}
\label{app:stellar}

\begin{figure}
\vspace{-1.4cm}
\centering
\begin{subfigure}{.48\textwidth}
  \centering
  \includegraphics[width=0.97\linewidth]{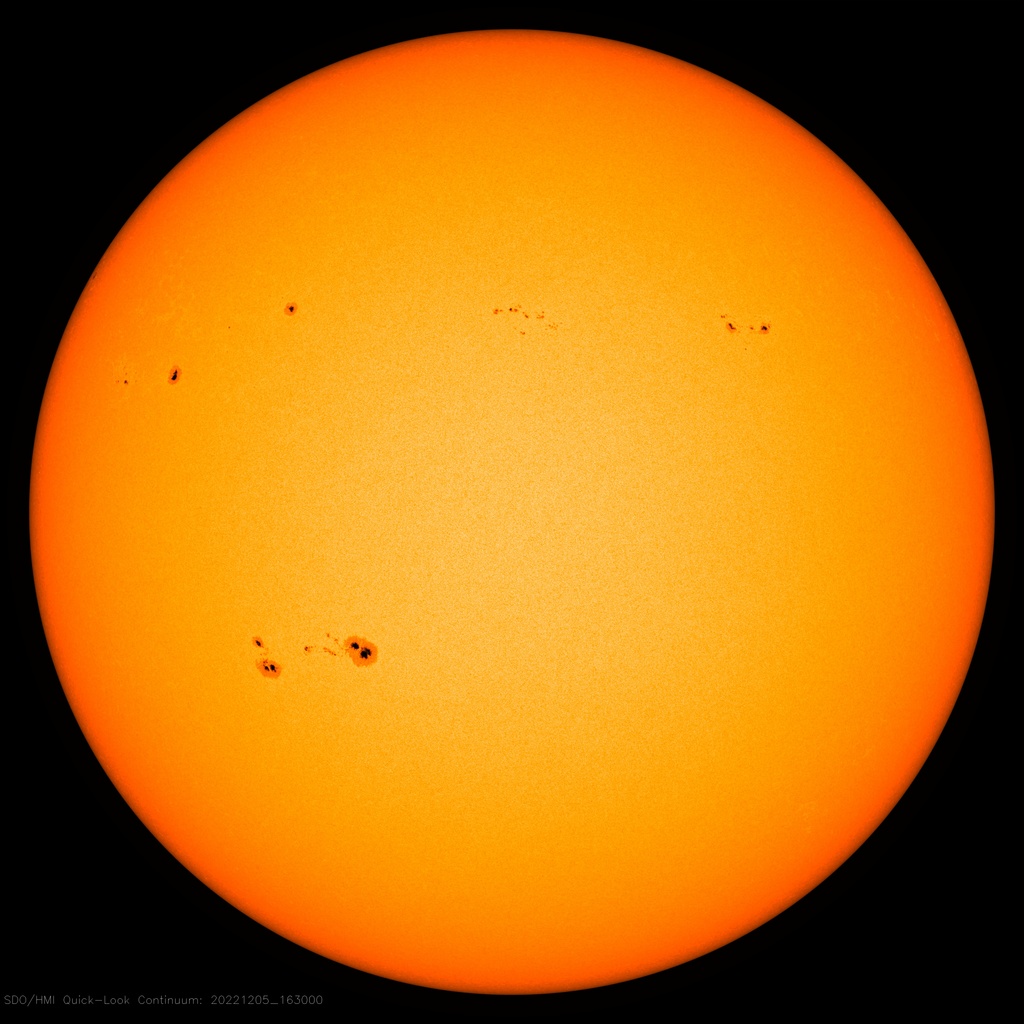}
  \caption{Full surface of the Sun, sunspots are the regions that appear darker than the continuum. The solar surface appears darker on the limb, a fact known as the limb-darkening effect.}
  \label{fig:sub1}
\end{subfigure}%
\hspace{0.3cm}
\begin{subfigure}{.48\textwidth}
  \centering
  \includegraphics[width=.96\linewidth]{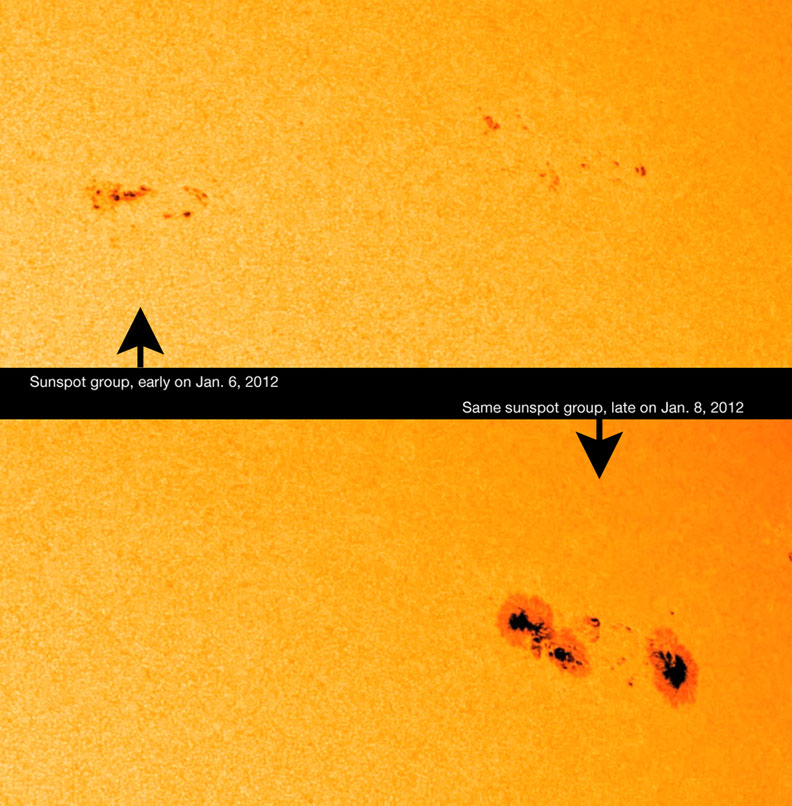}
  \caption{Same group of sunspots observed at two different dates, top: Jan 6 2012, bottom: Jan 8 2012.The granular aspect of the continuum comes from the convective motion of the gas at the stellar surface.  }
  \label{fig:sub2}
\end{subfigure}
\vspace{0.3cm}
\begin{subfigure}{.48\textwidth}
  \centering
  \includegraphics[width=1\linewidth]{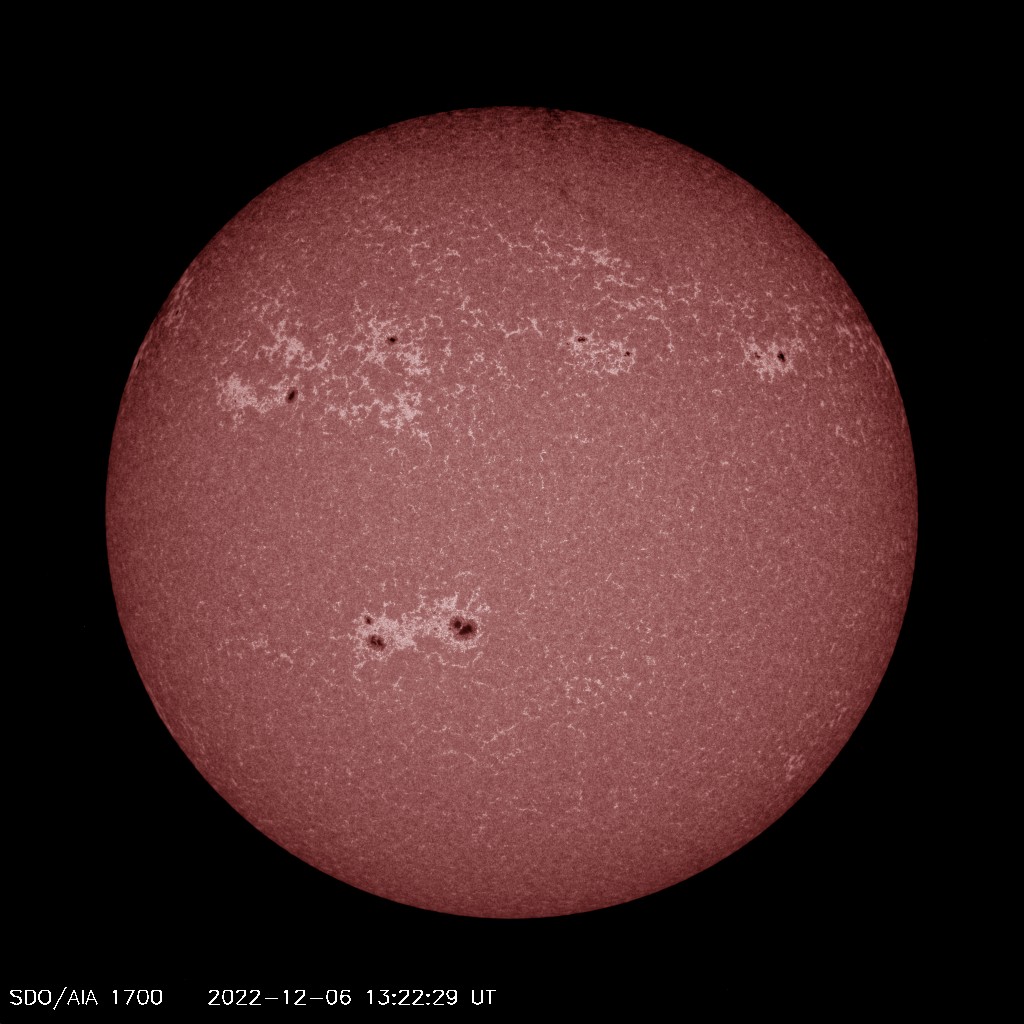}
  \caption{SDO observation of the Sun at wavelength 1700 \AA. The bright regions are called faculae, notice that faculae areas surround the spots (darker regions). Both faculae and spots inhibit the upward convective motion of the gas. Faculae cover more area than spots, but have a smaller temperature contrast with the continuum. The latitude, number and lifetime of the spots varies along stellar magnetic cycles, on the timescale of a few years (11 years for the Sun). }
  \label{fig:sub3}
\end{subfigure}
\hspace{0.3cm}
\begin{subfigure}{.48\textwidth}
  \centering
  \includegraphics[width=1\linewidth]{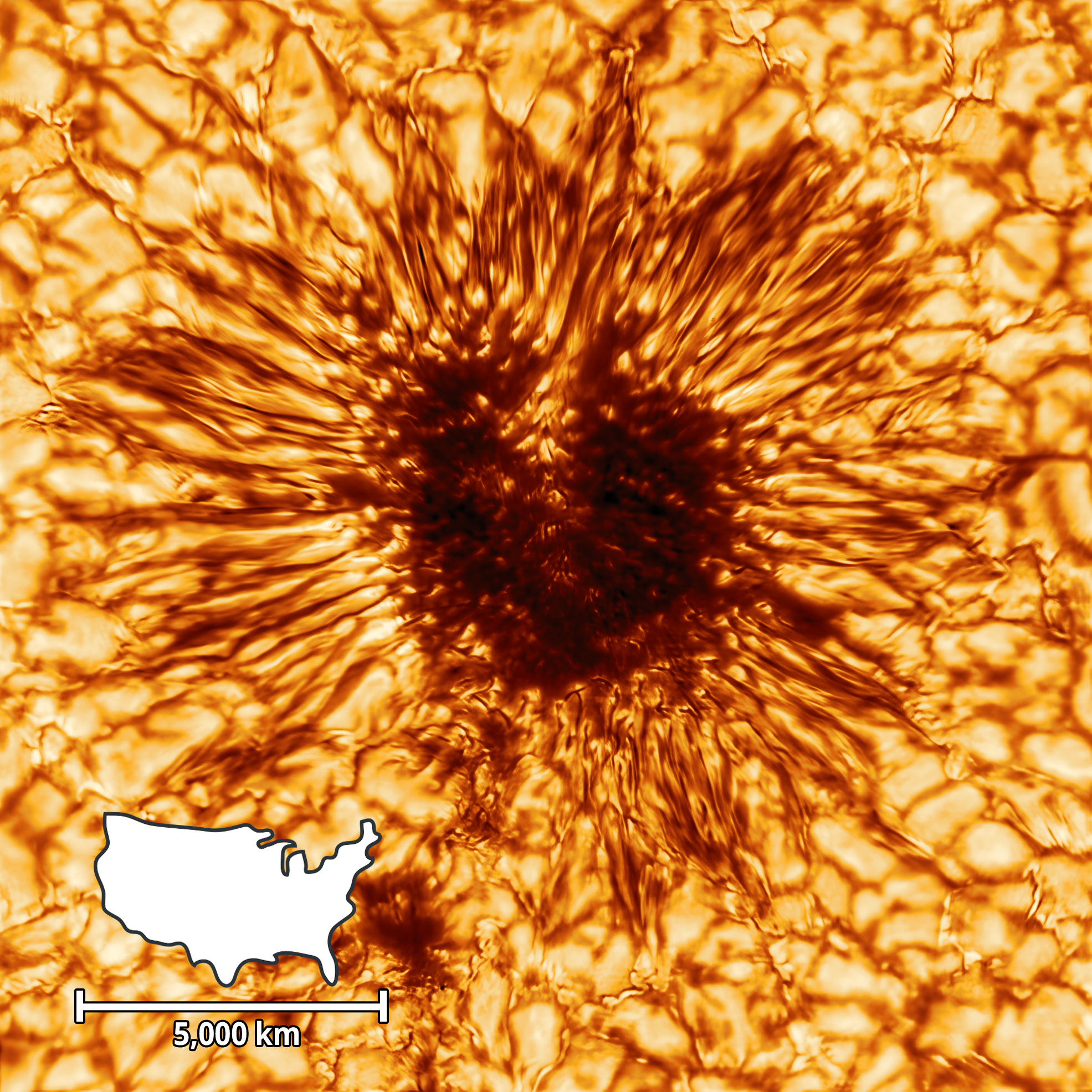}
  \caption{Inouye Solar Wave Front Correction (WFC) image, captured Jan 28, 2020, at 530 nanometers. The granulated structures around the spot are due to convection: hot plasma moves upwards at the center of granules, cools down and goes downwards between granules (darker inter-granular regions). The contribution of brighter intra-granular region exceeds that of intergranular ones, creating a so-called convective bluedhift on observed spectra.  Credit: NSO/AURA/NSF
 }
  \label{fig:sub4}
  \end{subfigure}
\vspace{0.4cm}
\caption{Solar dynamics observatory (SDO) images of the Sun in filtered light. Figures in (a), (b), and (c) are courtesy of NASA/SDO and the AIA, EVE, and HMI science teams.}
\label{fig:sdo}
\end{figure}

Understanding the impact of the variability of stellar surfaces on the signal is important to reach a precision below 1 m/s. There are several types of effects on stellar surface which affect the spectra measured and imprint onto the estimated radial velocities. 


 The physical processes taking place in stellar surfaces, how they vary depending on the type of star, and their impact on the data have been investigated through simulations, observations of the Sun and other stars. For a review of stellar effects affecting the RV, we refer the reader to \cite{cegla2019_review, meunier2021}.

Stars can be locally brighter and darker because of local excess magnetic field (see Fig. \ref{fig:sdo}). As the observed star rotates, the feature breaks the imbalance between the approaching and receding stellar limbs, respectively blue-shifted and red-shifted. Furthermore, these structures inhibit the upwards convective motion of the gas, an effect known as the convective blueshift inhibition \citep{dravins1981}. The effect of these magnetic structures on the radial velocity and activity indicators has been simulated for Sun-like stars in \cite{saar1997, hatzes2002, desort2007,boisse2012, dumusque2014, borgniet2015, dumusque2016_1,  meunier2019I, bellotti2021}, and for M dwarves in \cite{barnes2011}. Simulations can also be used directly to forward-model and analyse the data \citep{herrero2016}. The apparition of spots varies through the stellar magnetic cycle on the scale of $\sim$ 10 years, creating signals at this timescale \citep{makarov2010, meunier2013, fuhrmeister2022}.

Convection on the surface of the star creates a so-called granulation pattern (see Fig. \ref{fig:sub4}, and \cite{cegla2019_review} for a review). The effect of the changing patterns of convection on radial velocity and indicators changes with stellar spectral types \citep{meunier2017} and has been simulated in several works, \cite{cegla2013,meunier2015, cegla2018, dravins2021}. In simulations, the effect of this process on RV appears to be correlated to its effect on activity indicators \citep{meunier2013, cegla2019}, a fact that is still to be fully leveraged for the correction of the  RV granulation effect.


Stellar variability models need to be calibrated. This was initially done through the observation with the HARPS spectrograph of the solar light scattered by the asteroid 4/Vesta, with simultaneous high resolution observations of the solar surface with the solar dynamics observatory (SDO). From the surface observations, and based on the physical understanding of RV effect, the expected RV is computed and compared to the HARPS RV measurements \citep{haywood2016}. The lifetime and properties of granules on the Sun can be also derived from observations of the Sun \citep[e.g.][]{hirzberger1999}. An interesting observation is that, even though the lifetime of granules is $\sim 15$ min, averaging the solar surface for 1h does not completely average out their effect.  Furthermore, in recent years, several telescopes were built to observe the Sun and feed high precision spectrographs used to detect exoplanets at night. Such telescopes exist for HARPS-N \citep{dumusque2015solar, phillips2016}, 
HARPS and NIRPS \footnote{The HELIOS telescope \url{https://www.eso.org/public/usa/announcements/ann18033/#1}} spectrograph, NEID  \citep{lin2022}, EXPRES \citep{llama2022}, and Keck Planet Finder (KPF) \footnote{\url{https://www.caltech.edu/about/news/keck-observatorys-newest-planet-hunter-puts-its-eye-on-the-sky}}.
Several years of high-resolution solar spectra are publically avaliable from HARPS-N \citep{dumusque2021} and 
NEID \footnote{\url{https://neid.ipac.caltech.edu/search_solar.php}}.
And astronomers are beginning to 
draw conclusions about the Sun's variability based on Sun-as-a-star observations
\citep[e.g.,][]{colliercameron2019, miklos2020,almoulla2022}

Another well explored area consists in studying the spectroscopic observations of other stars, in particular the radial velocities and activity indicators. There has been substantial work on the observation of M dwarves \citep[e. g.][]{bonfils2007, forveille2009, suarezmascareno2018, zechmeister2018, jeffers2022}, in particular the periods of the exhibited variability, using the CARMENES sample \citep{schofer2019, lafarga2020, schofer2022}.
It was shown in particular that some activity indicators exhibit a closed-loop pattern. It means that  when plotting the time series of RV acquired at times $t_i, i=1,...,N$ not against time, but as a function of an activity indicator  at time $t_i$, the figure exibits a loop pattern. The behaviour of the indicators in Sun-like stars has been studied in \cite{gomesdasilva2011,boisse2011b, gomesdasilva2012, figueira2013, santerne2015, lanza2018, suarezmascareno2020}. The effect of granulation on RV and photometry has been studied in \cite{sulis2022} with simultaneous CHEOPS and ESPRESSO observations of two stars, of spectral type G and F.

More recently, attention has been given to the color-dependent effect of stellar activity: depending on their wavelength, spectral lines are affected differently \citep{thompson2017, wise2018, cretignier2020_lbl, ning2019, wise2022, almoulla2022a, lafarga2023}. This aspect will be crucial to leverage all the information present in the spectrum to estimate the contaminating signal in RVs (see Eq. \eqref{eq:rv_contribution_app}). This knowledge might be coupled to data-driven approach leveraging all the spetral information \citep{bedell2019, gullysantiago2022,gilbertson2023}.

In addition to magnetic activity and granulation, acoustic oscillations of the stars also affect the data \citep{dumusque2011i, chaplin2019, guo2022}. The most important oscillations for solar-type stars  (known as $p$ modes) have characteristic periods of a few to several minutes.  
Stellar Meridional winds \citep{beckers2007, makarov2010}, as well as gravitational redshift \citep{cegla2012} also affect the data at the level of a few cm/s.

To conclude this section, let us note that the problem laid out in \S\ref{sec:definitions} has some similarities with the setting of a technique called Doppler imaging, which relies on the following principle. As we have seen above, the presence of magnetically active region on the surface of the star changes the shape of the spectral lines. Conversely, one can infer the position and size of magnetic regions from the change of shape of the spectrum (for stars that rotate sufficiently fast). The inference is usually done on a weighted average of the spectral lines \citep{khokhlova1976, goncharskii1977, goncharskii1982, vogt1983, vogt1987}. From spectrapolarimetric observations, one can also retrieve some properties of the magnetic field \citep{donati1997}. In \S\ref{sec:definitions}, we do not aim at retrieving the distribution of stellar spots from a shape change in the spectrum, but at estimating the contaminating RV signal. Another approach consists of estimating the distribution of stellar spots with a Doppler imaging approach, and predicting the corresponding RV signal \cite{lanza2011, petit2015, barnes2017, yu2019, klein2022}, see \cite{barnes2015,barnes2017, hebrard2016, baroch2020} for the particular case of M dwarves. While this can work well for stars with one or two large active regions, the problem becomes underconstrained once one allows for several active regions being active at once and/or smaller active regions that evolve on shorter timescales.
\cite{luger2021_III} builds explicitly a likelihood for the Doppler imaging problem and provides a framework for marginalizing over surface maps.  While this can work well for stars with large active regions, applying it to active regions more typical of sun-like stars has yet to be proven practical, as it would require a very high-order spherical harmonic expansion of the solar surface.  


 



\section{PERIODOGRAMS}
\label{app:periodogram}

Let us suppose we have a timeseries with $N$ entries, $\vec y = (y(t_i))_{i=1,..N}$. We want to determine if $\vec y$ exhibits periodic variations, here potentially due to planets.
A common way to do this is to use a family of methods called periodograms. Such methods consist in computing the difference of log-likelihoods of a null hypothesis model $H_0$ and an alternative model $K_\omega$, including $H_0$ and one or several periodic signals. 
The simplest form of periodogram is the Lomb-Scargle periodogram \citep{lomb1976, scargle}. With the notations of Eq.~\eqref{eq:likelihood_simple_app} the alternative models are both Gaussian multivariate distributions such that 
\begin{align}
H_0: \; &\vec g  = \vec{0}, \label{eq:h0_app}\\ 
K_\omega :\; & \vec g(A,B,\omega)  = A\cos \omega   \vec t + B\sin \omega \vec t,  \label{eq:komega_app}
\end{align}
where $A$ and $B$ are chosen to maximize the log likelihood of $K_\omega$, which is here equivalent to finding $A$ and $B$ minimizing the sum of squares $\sum_{i=1}^N (y(t_i)-A\cos \omega (t_i) - B\sin \omega (t_i))^2 / \sigma_i^2$, where $\sigma_i$ is the error on measurement $i$. The null hypothesis expresses how the data is expected to behave in the absence of periodic components. For the Lomb-Scargle periodogram, the null hypothesis is that the data is a realization of a Gaussian, uncorrelated noise.  

In general, denoting by $\utheta_K$ and $\utheta_H$ the parameters of models  $K_\omega$ and $H_0$, the periodogram of time series $\vec y$ evaluated in frequency $\omega$ is
\begin{align}
\mathcal{P}(\omega) = \max_{\utheta_{K_\omega}} \log p(\vec y \mid \utheta_{K_\omega}) - \max_{\utheta_{H_0}} \log p(\vec y \mid \utheta_{H_0}) 
\end{align}
Other definitions are possible, in particular some authors normalize the periodogram by $\max_{\utheta_{H_0}} \log p(\vec y \mid \utheta_{H_0})$ \citep{baluev2008}.
The periodogram is computed on a grid of frequencies, $\Omega = (\omega_i)_{i=1..N}$.
We compute the maximum of the periodogram of the data to be analysed $\vec y$, which we denote by $\mathcal{P}_{\mathrm{d}} $, 
and define a false alarm probability, 
\begin{align}
    p(\max_{\omega \in \Omega} \mathcal{P}(\omega) \geqslant  \mathcal{P}_{\mathrm{d}} \mid H_0)
\end{align}
that is the probability that the maximum of the periodogram exceeds $\mathcal{P}_{\mathrm{d}}$ knowing $H_0$ is true. A small FAP means the hypothesis $H_0$ is unlikely. If one generates $M$ signals following the hypothesis $H_0$ and computes their periodogram, the FAP is simply the number of times the maximum of the periodogram exceeds $\mathcal{P}_d$ divided by $M$. This brute force approach gives a precision of 100/$M$ \% on the FAP. To evaluate whether the FAP is below 0.01 \% requires at least 10,000 simulations. However, thanks to the theory of extreme values of stochastic process, it is possible to obtain analytical approximations of the FAP associated to a certain definition of the periodograms, which greatly facilitates the use of periodograms \citep{baluev2008, baluev2009, baluev2013, baluev2013_vonmises, baluev2015, delisle2019a}. In the following, we will make explicit whether such a formula exists. In between the brute force and analytical approaches, the method of \cite{suveges2014} consists in generating a few datasets following the null hypothesis, and fit to the cumulative distribution function of maxima of the periodograms a generalised extreme value (GEV) distribution.

The periodogram has been extended to definitions of the likelihoods of $H_0$ and $K_\omega$ more complex than the Lomb-Scargle periodogram.The models are often assumed to be Gaussian, and each have a likelihood in the form of~\eqref{eq:likelihood_simple_app} 
In \citet{ferrazmello1981,cumming1999, reegen2007, zechmeister2009} the Lomb-Scargle periodogram is generalized by fitting for $c_0$, as defined in Eq.~\eqref{eq:nominal_model1_app} at each trial period.  \cite{ford2008} extends the definition of $H_0$ to $c_0\vec{1} + c_1 \vec t$ where $\vec t =(t_i)_{i=1..N} $.
These are particular instances of the case considered in \cite{baluev2008}, where $H_0$ is a linear model $\mat M \vec x$, $\mat M$ is a $N\times p$ matrix and $\vec x$ a vector of size $p$, 
 \begin{align}
H_0: \; &\vec g(\vec x)  = \mat M \vec x, \label{eq:h0_app2}\\ 
K_\omega :\; & \vec g(A,B,\omega, \vec x)  = \mat M \vec x  + A\cos \omega  \label{eq:komega_app2} \vec t + B\sin \omega \vec t, 
\end{align}
 where $\vec x, A$ and $B$ are fitted for each trial frequency to maximize the likelihood of model $K_\omega$. This framework can be applied to fitting separate offsets $c_0$ for each instrument. If there are $p$ instruments, one simply has to define $\mat M_{ij} = 1$ if measurement at time $t_i$ is taken by instrument $j$, and $\mat M_{ij} =0 $ otherwise.  This framework also allows to approximately compare $n+1$ vs. $n$ planets instead of 1 vs. 0, when including in the columns of $\mat M$ the $\cos \omega_i \vec t$ and $\sin \omega_i \vec t$ of the frequencies of planets found so far $\omega_i$, $i=1,..,n$.  \cite{baluev2008} provides an accurate analytical false alarm probability. 
 
 So far, $\mat V$ is supposed fixed. The noise model can also be generalized by allowing for an unknown $\sigma_J$ fitted at each frequency \citep{baluev2009}. Here, the noise is still assumed to be uncorrelated temporally, which is unrealistic in many cases due to the presence of stellar activity and instrumental systematics. The analytical FAP formula is generalised in~\cite{delisle2019a} to arbitrary, fixed covariance matrix $\vec V$, thus generalising the formula of  \cite{sulis2016} to the case of uneven sampling and non stationary Gaussian noises. It is also possible to fit the parameters of a non white noise model at each trial frequency, but this is costly and there is no analytical formula for the FAP \citep{delisle2018}. 
 
 Another line of work generalizes the form of the periodic function. Instead of fitting a sinusoid, the $K_\omega$ model can be defined as a Keplerian model \citep{cumming2004, zechmeister2009}. This is particularly useful to detect signals of planets with high orbital eccentricity. \cite{baluev2015} provides a rigorous analytical FAP in that case, as a special case of analytical FAPs for non-sinusoidal signals \cite{baluev2013_vonmises}. 

 Furthermore, instead of searching for one periodicity, it is possible to search for multiple planets simultaneously \citep{ford2011,baluev2013}, and jointly modeling RVs and other stellar ancillary indicators \citep{delisle2022,jones2022}.
 
 Instead of computing the maximum likelihood models for $H_0$ and $K_\omega$, one can marginalize over certain parameters. It means that instead of maximizing $p(\vec y \mid \utheta_{K_\omega})$ over $\utheta_{K_\omega}$, we separate the parameters in two categories $\utheta_{K_\omega} = (\utheta_{K_\omega}^1, \utheta^2 )$, $\utheta_{H_0} = (\utheta_{H_0}^1, \utheta^2 )$, and compute
 \begin{align}
     p( \vec y \mid \utheta_{K_\omega}^1) = \int  p( \vec y \mid \utheta_{K_\omega}^1,\utheta^2 ) p(\utheta^2) \dd \utheta^2 \;\; ;  \;\;  p( \vec y \mid \utheta_{H_0}^1) = \int  p( \vec y \mid \utheta_{H_0}^1,\utheta^2 ) p(\utheta^2) \dd \utheta^2
 \end{align}
Typically $\utheta_{K_\omega}^1$ is the frequency and $\utheta^2 $ are parameters on which the model depends linearly. In that case, marginalization can be done with analytical expressions \citep{cumming2004, mortier2015, feng2017}. Marginalizing can also be done via the Laplace approximation \citep{ford2008, nelson2020} or by using numerical quadrature. The latter method has been used to marginalize over a noise term $\sigma_J$ at each trial frequency \citep[][Appendix A3]{nelson2020}. Conceptual and practical differences between using the maximization or marginalization of the likelihood are discussed in~\cite{baluev2022}. 

The type of periodogram to use depends on the trade-off between precision and speed needed. The availability of an accurate analytical FAP is the major driver of the speed, since there is no need to compute periodograms several times. Analytical estimates listed above are reliable for signals with a FAP of 0.1 or lower. All periodograms involving a linear fit with an analytical FAP~\citep{baluev2008, delisle2019a} are very fast (a few seconds on a typical dataset, on a laptop). Keplerian or multi frequency periodograms can take longer (several minutes at least).

\section{BAYESIAN MODEL COMPARISON: COMPUTATIONAL CHALLENGES}
\label{sec:computations}

Bayesian model comparison involves computing the Bayesian evidence, of marginal likelihood. The Bayesian of evidence (also called marginal likelihood) of a $n$ planet model is
\begin{equation}
    p(\vec y \mid n) = \iint_{\Theta_n} p(\vec y \mid \utheta, \ueta, n ) p(\utheta, \ueta\mid n) \dd \utheta \dd \ueta . 
    \label{eq:evidence_app}
\end{equation}
and the ratio $p(\vec y \mid n+1)/ p(\vec y \mid n)$, referred to as a Bayes factor, is a very common statistical significance metric for exoplanet detection. If $n$ planets have been confidently detected, and if the Bayes factor is above a certain threshold, the detection of the $n+1$th planet is claimed. The computation of the evidence and a reliable estimation of the posterior distribution $p( \utheta, \ueta \mid \vec y, n )$ are both crucial for a useful Bayesian detection criteria that results in the posterior probability for a planet to be within a confined region of the parameter space \citep{brewer2015, faria2022, hara2021a}.

Evaluating an integral over a large parameter space is notoriously difficult. One of the difficulties of radial velocity data is that the model contains periodic signals. The frequency resolution is approximately 2$\pi/T_{\mathrm{obs}}$ where $T_{\mathrm{obs}}$ is the total timespan of observation. As a result, in a frequency range $[\omega_1, \omega_2]$ with $\Omega = \omega_2 - \omega_1$, the number of local likelihood maxima  is approximately $N_{\mathrm{locmax}} = \Omega T_{\mathrm{obs}}/ 2 \pi $. If there are $k$ periodic signals with frequencies in the same range, we should expect $\sim N_{\mathrm{locmax}}^k$ local minima when varying the frequencies. Furthermore, the geometry of the Keplerian signals is such that there are many local minima at high eccentricity \citep{baluev2015, hara2019ecc}.

To compute Eq.~\eqref{eq:evidence_app}, two family of methods are used: Monte-Carlo Markov Chain (MCMC) and nested sampling. In both cases, special attention must be paid to the exploration of the complex parameter space. MCMC methods rely on the definition of a proposal distribution, which must lead to samples both jumping from local minima far from each other, and explore the dominant modes. The numerical difficulty of nested sampling algorithms is to find efficiently a sequence of independent samples following the prior distribution restricted to regions of the parameter space with sufficient likelihood.

\citet{fordgregory2007} compared several methods for estimating the Bayesian evidence, and found that several performed particularly poorly for our standard model.
Methods that performed well for the 1-planet model were: the Laplace approximation, importance sampling, the ``ratio estimator'' \citep{nelson2016}, and parallel tempering Monte-Carlo Markov Chain (MCMC) \citep{gregory2007}.  
Subsequently, several authors have applied nested sampling algorithms~\citep{brewer2014, handley2015, faria2016, buchner2021}, which couple the estimation of the posterior distribution with the Bayesian evidence (Eq.~\ref{eq:evidence_app}). 

In order to  better understand the strengths and limitations of approaches for computing Bayesian evidences, \citet{nelson2020} designed a data challenge in coordination with the 3rd Extreme Precision Radial Velocity workshop (Penn state, August 14-17 2017).  
Nine participants/teams were given two sets of priors, and four likelihoods (corresponding to 0, 1, 2 and 3-planet models) and 6 simulated datasets.  
Teams compared their results for the first dataset during the workshop, refined their methods and submitted updated results.  
Collectively, teams reported estimates of the marginalized likelihood and its uncertainty using fifteen computational methods.

As expected, some methods (e.g., Bayes Information Criterion, Chib's estimator) gave highly discrepant results for Bayes factors. Laplace approximation typically agreed with other high-quality methods to within a factor of $\sim~10^2$ for 0 to 2-planet models and $\sim~10^4$ for a 3-planet model.
Of the remaining methods, there was general agreement for the Bayes factor to within a factor of $\sim~10$ for 1-planet models, and $\sim~10^2$ for 2 and 3-planet models. 
Methods that demonstrated broad agreement included: 
combining MCMC with importance sampling using either the Perrakis estimator \citep{perrakis2014} or the ratio estimator \citep{nelson2016}, variational Bayes with importance sampling \citep{beaujeancaldwell2013}, and multiple forms of nested sampling \citep{veitchvecchio2010,feroz2009,feroz2013}.
It is unclear why diffusive nested sampling \cite[using DNest4;][]{brewer2016} sometimes gave results significantly discrepant from the other Bayesian methods.  

The significant dispersion in numerical estimates of Bayes factors emphasizes the importance of adopting a threshold Bayes factor much larger than is typically recommended in the literature (often 150, in the context of much simpler models).  
At first, it may seem concerning that different estimates of Bayes factors could vary by a few orders of magnitude.  
In practice, Bayes factors for RV datasets are often so extreme that an uncertainty of ``only'' a few orders of magnitude is acceptable.
Of course, publication biases can result in planet discoveries being preferentially reported when the Bayes factor is still ``only'' a few to several of orders of magnitude from unity.  

Another important result of this data challenge was that the observed dispersion of Bayes factors estimates was often significantly greater than predicted for some methods. Therefore, when using iterative algorithms, it is important to compute and report uncertainties for Bayes factors based on multiple independent runs.

\section{MODEL: SPECIFYING THE PRIORS AND LIKELIHOODS}
\label{sec:GP_activity_app}

\subsection{Priors}
\label{sec:bayesian_priors}

The choice of priors affects both orbital parameter estimates and model comparison to determine the number of planets, especially for low amplitude signals, where the likelihood is flatter \citep{hara2021a}.   
Based on a combination of astrophysical and statistical considerations, \citet{fordgregory2007} recommend the following priors:
$p(\log P) \sim U[\log P_\mathrm{min}, \log P_\mathrm{max}]$
$p(e) \sim U[0,1)$, 
$p(\omega) \sim U[0,2\pi)$, 
$p(M_0) \sim U[0,2\pi)$, and
$p(c_0) \sim U[-c_{0,\mathrm{max}},c_{0,\mathrm{max}}]$, 
where $U$ represents a uniform prior with the given bounds.
For minimally informed priors of scale variables, 
\citet{fordgregory2007} introduce a prior distribution \ch{for a real variable $z \in [0, z_{\mathrm{max}}]$, defining $0<z_0,z_{\mathrm{max}}$}
\begin{equation}
    p_{FG}(z \mid z_o, z_{\mathrm{max}} ) = \frac{(z_o+z)^{-1}}{\log(1+z_{\mathrm{max}}/z_o)},
\end{equation}
they recommend $p(K \mid P) \sim p_{FG}(K\mid K_o, K_{\mathrm{max}} (P_{\mathrm{min}}/P)^{1/3})$ and $p(\sigma_J) \sim p_{FG}(\sigma_J \mid\sigma_{J,o}, K_{\mathrm{max}})$.  
Physically motivated choices for $P_{\mathrm{min}}$ and $K_{\mathrm{max}}$ vary with the target star properties. \ch{For instance, the values of $K$ and $P$ corresponding to a planet that would be disintegrated with tidal forces, that is beyond the Roche limit, are excluded.} 
Typical values for a sun-like star would be $P_{\mathrm{min}} = 0.2$ days and $K_{\mathrm{max}} = 3.6$ km/s.   
Non-zero values of $K_o$ and $\sigma_{J,o}$ are necessary to ensure proper priors.  
\citet{fordgregory2007} recommend $K_o = \sigma_{J,o} = \sigma_{\mathrm{RV},o} \sqrt{50/N_{\mathrm{obs}}}$, where $\sigma_{\mathrm{RV},o}$ is the typical measurement precision and 
$N_{\mathrm{obs}}$ is the number of observations, so as to provide sensitivity to planets within the realm of detection without overly penalizing models for planets that are essentially impossible to detect with available data.  
%
%
%
Posterior distributions for $e$ and $\omega$ vary considerably with the quality of the data. As geometric considerations dictate the prior for $\omega$, the only priors that are likely to affect the estimation of orbital parameters are those of $K$ and $e$.

\begin{table}
\caption{Reference Priors\label{tab:priors_common}}
\begin{center}
\begin{tabular}{@{}lcccc@{}}
\hline
Parameter & Variable & Unit & Distribution & Note \\
\hline
Orbital Period & $P$ & d & $\log P \sim U[\log P_{\min}, \log P_{\max}]$  & $a,b$   \\\hline
Velocity Amplitude & $K$ & m s$^{-1}$ & $\log K \sim p_{FG}(K_o, K_{\max} \left(\frac{P_{\min}}{P}\right)^{1/3})$ & $c,d$   \\\hline
Orbital Eccentricity & $e$ & \ldots & $e \sim U[0, e_{\max}(P))$ & $e$   \\\hline
Argument of Pericenter & $\omega$ & radians & $\omega \sim U[0,2\pi)$  & f   \\\hline
Mean Anomaly at epoch & $M_0$ & radians & $M_0 \sim U[0,2\pi)$  & f   \\\hline
RV offset & $c_0$ & m s$^{-1}$ & $c_0 \sim N[0,\sigma^2_{\mathrm{disp}}]$  & g   \\\hline
RV slope & $c_1$ & m s$^{-1}$ yr$^{-1}$  &  $c_1 \sim N[0,\sigma^2_{c_1}] $  & h   \\\hline
\hline
\end{tabular}
\end{center}
\begin{tabnote}
$^{\rm a}$ For sun-like stars, astrophysical considerations justify hard limits of $P_{\min}\simeq~0.2$d and $P_{\max}\simeq~10^8$ year.  However, theory predicts that the occurrence rate drops dramatically at much shorter orbital periods ($\simeq~1000^3$yr).  
In practice, RV surveys are limited by the survey duration and the choice of $P_{\max}$ is only important when performing model comparison.
$^{\rm b}$ Empirically, the planet occurrence rate decreases for periods less than $3$d.  For an analytical reference prior that limits how much very-short period planets are overweighted relative to nature, one might adopt $P\sim p_{FG}(3\mathrm{d}, P_{\max})$, truncate assign no probability to periods less than $P_{\min}$.
$^{\rm c}$ The intrinsic occurrence rate of small planets may to continue rising to smaller amplitudes than are typically detectable for the foreseeable future.  In practice, we recommend setting $K_o = \sigma_{RV,1\mathrm{obs}} \sqrt{50/N_{\mathrm{obs}}}$, so as to reflect the limiting precision of an RV survey, where $\sigma_{RV,1\mathrm{obs}}$ is measurement precision of a single observations and $N_{\mathrm{obs}}$ is the number of observations.  Since there will not be strong evidence for a planet with $K$ less than the detection limit of a survey, there is no benefit to assigning substantial prior mass to such planets.
$^{\rm d}$ For Sun-like stars and $P_{\min}=0.2$d, $K_{\max} \simeq~3.6$km/s.  Larger RV signals would be due to perturbations by a stellar binary companion, rather than an exoplanet. 
$^{\rm e}$ Keplerian orbital models develop pathologically narrow spikes for $e\simeq~1$.  Formally, one could avoid these by imposing a maximum eccentricity, $e_{\max}(P) = 1-\left(\frac{r_{p,\min}}{AU}\right)\left(\frac{P}{\mathrm{yr}}\right)^{-2/3}\left(\frac{M_\star}{M_\odot}\right)^{-1/3}$, so as to ensure the planet's pericenter distance exceeds the Roche limit, 
$r_{p,\min}$.  In practice, most MCMC and nested samplers do not find the pathological solutions for $e\simeq~1$ and thus practioners set $e_{\max}=1$, but ignore the contributions from $e\simeq~1$ solutions. \cite{kipping2014} uses a beta prior fitted on the eccentricity of known giant planets, which disfavours high eccentricities.
$^{\rm f}$ Geometric considerations provide a strong motivation for this choice.
$^{\rm g}$ For surveys of sun-like stars in the solar neighborhood, a dispersion of $\sigma_{\mathrm{disp}}\simeq~40$km s$^{-1}$ would be well-motivated by knowledge of Milky Way structure.  While astrophysical knowledge could motivate alternative choices for other populations, in practice this has minimal effect as long as the prior assigns non-zero mass to the true mean radial velocity offset.  In practice, most practitioners assign a uniform prior.
$^{\rm h}$ $\sigma_{c_1}\simeq~10$ km s$^{-1}$ is representative of the acceleration due to a wide binary companion.  The specific choice of prior is unlikely to effect parameters for any well-characterized planets once the timespan of observations exceeds two orbital periods.  
\end{tabnote}
\end{table}

The choices made so far consider that the different orbital elements are \ci{statistically} independent. However, some combinations of parameters are less likely than other. For instance \cite{steffen2010} uses a Rayleigh prior on eccentricity stemming from dynamical scattering models in the solar system. In multiplanetary systems, certain parameter configurations lead to dynamically unstable systems and can be ruled out. Ensuring so-called AMD stability, which is necessary for the system to be stable \citep{faria2022}, or using randomly selected samples to set the initial conditions of a $n$-body simulation of the system, and simulations that end up violating a stability condition are dropped \citep{hara2020, stalport2022}. To speed up computations, the stability for a given set of orbital elements can be predicted with a trained neural network \citep{tamayo2020}. 

While the \ch{uncertainties on orbital elements obtained from} posterior distribution for RV-detected exoplanets are usually insensitive to the priors, those for $P$, $K$ and $\sigma_J$ can significantly affect the marginalized likelihood and thus can influence Bayesian model comparison. Indeed, the broader the priors, the more costly it is to add a planet. The effect of $P$ and $K$ priors is discussed in \cite{hara2021a}.


One could attempt to update the priors above based on the substantial improvements in our understanding of the distribution of exoplanets over the past decade \citep[e.g.,][]{he2021_amd, kipping2014}. \ch{This is particularly difficult in RV, subject to complex observational biases due to frequent human intervention in observational strategies. Thus, at present we recommend that standard practice use the analytic priors such as those above, potentially refined by a stability analysis}.  

In our nominal model \ch{(see Eq.~\eqref{eq:nominal_model1}), we add a trend} $c_1$, for which we recommend,
$p(c_1) \sim N(0, \sigma_{c_1}^2)$, with $\sigma_{c_1} = 10$ m/s/yr, representative of an acceleration due to a wide binary companion.   Posteriors for $c_0$, $c_1$ and $\sigma_J$ are typically well-constrained, but of secondary interest. The priors on the correlated noise model must be chosen carefully and with relatively strict bounds, otherwise the noise model can absorb viable planetary signals.



\subsection{Contaminating signals as Correlated Noise} 
\label{sec:stellar_variability_as_correlated_noise}
%
Signals induced by stellar activity and instrumental systematics can be represented as Gaussian noises with vanishing mean and a certain covariance function conveying their self similarity as a function of time. Below, we focus on the representation of stellar noises, as there are no consensus on stochastic representations for systematics. However the models adopted can also absorb some of the instrumental signals.


\subsubsection{Granulation}
\label{sec:likelihood_granulation}

Section 2.4.1 of the main document introduced a type of stellar variability called granulation. The RV variations induced by this phenomenon are modelled as a random process with a super-Lorentzian profile, which has a power density spectrum in the form $P(\omega) = S_0/(1 + \omega^{a}/ \omega_0^{a})$ where $S_0,\omega_0$ and $a$ are free parameters \citep{harvey1985, dumusque2011i, cegla2018, guo2022}, or as a sum of such processes. 
A value of $a \sim 4$ seems to be favoured by photometric and RV observations \citep[][respectively]{kallinger2014, guo2022}, although there can be a higher discrepancy~\citep{dumusque2011i, cegla2018}. With this value, the covariance matrix $\mat V$ in Eq.~\ref{eq:likelihood_simple_app} can be written \citep{foremanmackey2017,luhn2022}
\begin{equation}
\mat V_{i,j}^{\mathrm{gran}}(S_0,\omega_0) = k^\mathrm{gran}(|t_i-t_j|) = 
\sum_{g=1}^{N_g} S_g \omega_g \e^{-\frac {\omega_g|t_i-t_j|}{\sqrt{2}}} 
\cos\left(\frac{ \omega_g|t_i-t_j|}{\sqrt{2}} - \frac{\pi}{4}\right),
\label{eq:kernel_exp}
\end{equation}
\ch{where $S_g$ and $\omega_g$ parametrize the amplitude and correlation time-scale.} Based on astreroseismic observations, astronomers typically invoke 2-4 granulation timescales, potentially ranging from minutes to days. Theoretical considerations also provide guidance for relationships between granulation amplitudes and timescales as a function of stellar properties \cite{guo2022}. \ch{ \cite{cegla2019} study the correlations between the RV and ancillary indicators time series and find in particular that the so-called line equivalent width and the total photometric flux correlate with RV, although the correlation reduces with an increased rotational velocity of the star.   }

\subsubsection{Magnetic activity}  \label{sec:magnetic_activity}
Next, we consider the effect of magnetically active regions (e.g., stellar spots and faculae).  Large active regions can persist for more than one rotation time, leading to a quasi-periodicity in the measured RVs, since each  active region grows, shrinks and changes in strength and shape over its lifetime. 
As stellar rotation periods are often comparable to plausible orbital periods, the quasi-periodic nature of active regions makes them particularly problematic when trying to detect and build confidence in a putative planet candidate. 
One kernel commonly used to model this behaviour is  \citep{aigrain2012, haywood2014}, 
\begin{equation}
   \mat V_{i,j}^\mathrm{QP}(\eta_1, \eta_2, \eta_3, \eta_4) = k^\mathrm{QP}(|t_i-t_j|) =\eta_1^2 \exp \left[- \frac{(t_k-t_l)^2}{2 \eta_2^2}  - \frac{2 \sin^2  \frac{\pi (t_k-t_l)}{\eta_3} }{\eta_4^2}\right] ,
   \label{eq:kernel_qp}
\end{equation}
is the product of a Gaussian decay with time-scale $\eta_2$ and a periodic function with period $\eta_3$ that is often associated with the stellar rotation period. The parameter $\eta_4$ controls how close the $\Delta t$ must be to a multiple of $\eta_3$ for there to be a strong periodicity. Low values of $\eta_4$ lead to a very peaked correlation around the multiple of $\eta_3$, which is unphysical, and $\eta_4 \in [0.5,2]$ is often used.

On Sun like stars, spots have a tendency to appear on diametrically opposed longitudes of the star \citep{borgniet2015}. Based on simulations, \cite{perger2021} suggest using 
\begin{eqnarray}
k^{\rm QPC}(\tau)  =  \exp \Big( -2\frac{ \tau^{2}}{\lambda^{2}} \Big) \cdot \Bigg[    h_1^2 \, \exp \Big(- \frac{1}{2 \, w_0^2} \, \sin^{2}{\left(\pi \frac{\tau}{P}\right)} \Big) + h_2^2 \, \cos{\left(4\pi \frac{\tau}{P}\right)}  \Bigg]. \nonumber\label{Ec3}
\label{eq:qpc}
\end{eqnarray}
where $ h_1,  h_2, \lambda$ and $P$ are the free parameters of the kernel representing the amplitude of the quasi periodic part of the kernel, of the first harmonic of rotation due to the presence of spots on opposed longitudes, the coherence timescale of the noise and the stellar rotation period. See \cite{nicholson2022} for a discussion of the QP and QPC kernel parameters and stellar properties.

\cite{foremanmackey2017} suggest to use the kernel corresponding to a stochastically excited harmonic oscillator, the SHO kernel,
\begin{equation}
k^\mathrm{SHO}(\tau;\,S_0,\,Q,\,\omega_0) =
S_0\,\omega_0\,Q\,e^{-\frac{\omega_0\,\tau}{2Q}}\,
\begin{cases}
    \cosh{(\eta\,\omega_0\,\tau)} +
        \frac{1}{2\,\eta\,Q}\,\sinh{(\eta\,\omega_0\,\tau)}, & 0 < Q < 1/2\\
    2\,(1+\omega_0\,\tau), & Q = 1/2\\
    \cos{(\eta\,\omega_0\,\tau)} +
        \frac{1}{2\,\eta\,Q} \sin{(\eta\,\omega_0\,\tau)},& 1/2 < Q\\
\end{cases}
\label{eq:sho}
\end{equation}
where $S_0, \omega_0$ and $Q$ are the free parameters of the kernel, describing its amplitude, natural frequency and quality factor, respectively. The higher the quality factor, the more the  power spectrum density (PSD) is peaked around the natural frequency. For low quality factors, the PSD is decreasing with frequency, and might represent granulation. The kernel \eqref{eq:kernel_exp} is a SHO kernel with $Q=1/\sqrt{2}$.
The covariance matrix corresponding to a SHO kernel has the advantage of being semi-separable, such that computing $\mat V^{-1} (\vec y - \vec g)$ (see Eq. \eqref{eq:likelihood_simple}) has a $O(N)$ cost. 

Quasi periodic kernel can be useful for stars with large, long-lived active regions, most active regions for relatively quiet stars like the Sun do not last a rotation period.  Based on modeling simulated solar time series, \citet{gilbertson2020} recommend \ch{the kernel}
\begin{equation}
   k^{a}(\tau) =\eta_1^2 k^{M_{5/2}}(\tau/\eta_3) - \eta_2 \frac{d^2}{dt^2} k^{M_{5/2}}(\tau/\eta_3),   \label{eq:kernel_ar}
\end{equation}
where $k^{M_{5/2}}$ is the Matérn-5/2 kernel,
\begin{equation}
   k^{M_{5/2}}(\tau) =  \sigma^2 \left( 1 + \frac{\sqrt{5}\tau}{\rho} + \frac{5\tau^2}{3\rho^2} \right) \e^{-\frac{\sqrt{5}\tau}{\rho}}
   \label{eq:matern}
\end{equation}
$\eta_3$ is a characteristic timescale, and the strength parameters $\eta_1 \simeq 0.1069$ and $\eta_2 \simeq 1.154$ were calibrated based on simulations active regions for the Sun. 
Other kernels in use include Gaussian, exponential or kernels based on the physics of the star \citep{baluev2018, luger2021_II}.
In Fig. \ref{fig:kernels}, we represent the kernels described above with different values of their parameters. 

\begin{figure}[h]
\includegraphics[width=\linewidth]{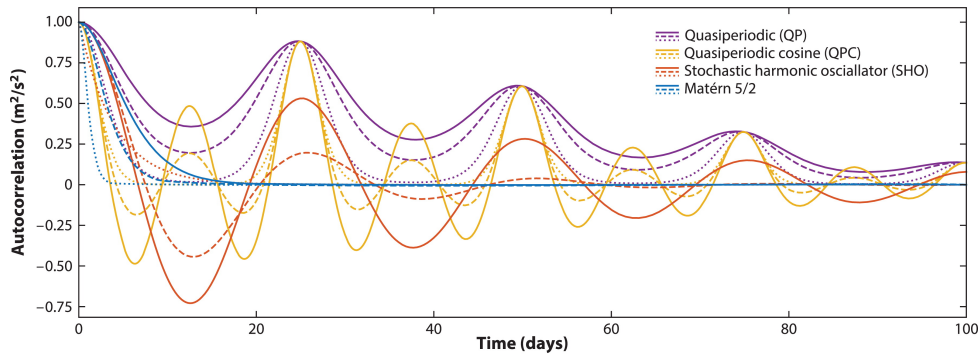}
\caption{Kernel functions: functions of the time lag $\Delta t$ between two observations, normalized so they are equal to 1 in $\Delta t = 0$. The covariance matrix $\mat V$ in Eq.~\eqref{eq:likelihood_simple_app} is such that its element $i,j$ is a sum of kernel functions evaluated in $\Delta t = t_i - t_j$. Each color corresponds to a different family of kernels: Quasi periodic (purple, Eq. \eqref{eq:kernel_qp}, with $\eta_1 = 1$ m/s, $\eta_2 = 50$ days, $\eta_3 = 25$ days, $\eta_4 = 0.5,1.25,2$ corresponding to dotted, dashed and plain lines respectively, Quasi periodic cosine (yellow, Eq. \eqref{eq:qpc}) with $\lambda_0 = 100 $ days,  $\omega_0 = 0.31$, $h_2/h_1 = 0,0.5,1$ corresponding to dotted, dashed and plain lines respectively, SHO (red, Eq. \eqref{eq:sho}) with $S_0=1$ m/s, $\omega_0=2\pi/25 $ rad/day, $Q=1/2, 2, 5$  corresponding to dotted, dashed and plain lines respectively, and Matèrn 5/2 (blue, Eq. \eqref{eq:matern}) with $\sigma =1$, $\rho=1, 3, 5$ days  corresponding to dotted, dashed and plain lines respectively. }
\label{fig:kernels_app}
\end{figure}

The analytical form of the kernels listed above rely on qualitative considerations and fits to simulations, but does not result from physical first principles. The approach taken by \cite{luger2021_III} expresses a physical linear transformation giving the spectrum as a function of the state of the stellar surface. It considers that the brightness on the surface of the star, decomposed in spherical harmonics, is a Gaussian process. Applying the linear transformation to the surface yields a two-dimensional Gaussian process representation of the time series of spectra. The framework of \cite{luger2021_III} is primarily thought for Doppler imaging: reconstructing the stellar surface from a time series of spectra \citep[e.g.][]{vogt1983, vogt1987}. \cite{hara2023} uses another way to derive kernels with physical interpretation of radial velocity, indicators and photometry -- either considered individually or jointly -- which take into account the varying properties of spots and faculae along the magnetic cycle. 


In the scientific literature dedicated to RV, kernels modelling stellar activity have been assumed to be functions of $|t'-t|$ \cite[except in][]{baluev2015_55cnc}),  
However, the kernel can depend not only on the time difference between measurements, but on their absolute values. \ch{Such non stationarity is observed for instance in instrumental calibration noise} \citep{delisle2019b}. 
Stationarity is not fully justified, since the magnetic cycles of stars modulate the strength of stellar activity from year-to-year 
(see section 2.4.1 of the main document), but is realistic over a few years.

\subsubsection{Oscillations}
\label{sec:likelihood_pmodes}

Stellar oscillations also contribute to the observed RV signal. 
For the Sun, the characteristic timescale for p-mode oscillation is $\sim~5$ minutes \ci{and is usually dealt with through observational strategies~\citep{dumusque2011i, chaplin2019}.}
 \ci{In the case of solar-type stars}, RVs come from a complex spectrum of $\sim~30$ p-modes, each with a lifetime of days.  
Astronomers often invoke with a continuous mixture of modes with a Gaussian envelope for their characteristic amplitudes. 
Such oscillations can be described by a stochastic harmonic oscillator kernel~\citep{foremanmackey2017}.
%
In practice, the exposure time is usually a non-negligible fraction of the characteristic timescale, so one should integrate the kernel over the exposure times; see \citet{luhn2022}.


\subsubsection{Computational Considerations}
\label{sec:correlated_noise_in_practice}

Evaluating the likelihood Eqn.~\ref{eq:likelihood_simple_app} for certain values of the parameters requires the log determinant of $\mat V(\ueta)$ and computing $\{\vec y - \vec  g(\vec t; \utheta)\}^T \mat V^{-1}(\ueta) \{\vec y - \vec g(\vec t; \utheta)\} $.  With standard algorithms the computational cost scales as $O(N^3)$ with the size $N$ of the dataset $\vec y$, while more advanced or approximate algorithms can reduce this considerably.  As seen in \S\ref{sec:computations}, evaluating the posterior distribution and Bayesian evidence requires millions of likelihood evaluations, and makes the analysis of large datasets computationally challenging. Fortunately, the covariance matrix is semi-separable for certain kernels and then the computational cost scales as $O(N)$. \cite{foremanmackey2017} present the \texttt{celerite} framework\footnote{\url{https://github.com/exoplanet-dev/celerite2}}, 
which models covariances as sums of constant, cosine and sine functions. They suggest modeling stellar activity with the covariance of a stochastically excited harmonic oscillator. 
Similarly, \texttt{TemporalGPs.jl} \footnote{\url{https://github.com/JuliaGaussianProcesses/TemporalGPs.jl}} allows for $O(N)$ computation of a broad class of 1-d kernel functions in Julia.  
The CELERITE framework is generalised by~\cite{delisle2019b} to the S+LEAF framework, which models covariances as a sum of semi-separable and so-called LEAF matrices\footnote{\url{https://gitlab.unige.ch/Jean-Baptiste.Delisle/spleaf}}, useful to model the noise due to the calibration of the instrument.
 Quasi-separable kernels provide even greater flexibility for modeling multivariable timeseries (\texttt{tinygp} \footnote{\url{https://tinygp.readthedocs.io/en/stable/index.html}}).




\subsection{Incorporating Ancillary Time Series With Gaussian Processes} 
\label{app:multidim_gp}

Due to the high flexibility of correlated noise models, true RV variations due to planets can often be explained as correlated noise.
In principle, one can compare the marginalized likelihood of models with no, one or multiple planets using a correlated noise model, but when the parameters of the noise model of Eq.~\eqref{eq:kernel_qp} are treated as free parameters in the computation of the Bayesian evidence, this often leads to overfitting that increases the likelihoods and decreases the evidence for planets, including those that are actually present. 
Hence, it is highly desirable to incorporate more information in order to constrain the stellar activity model more tightly.
So far, we have assumed that the measured RVs are analyzed separately from any ancillary time series (e.g., other summary statistics computed from the spectra or photometry observations), but, as argued in  \S2.4.1. of the main document, features on the surface of the star should affect other properties of the spectrum.  Jointly analysing a multivariate time series constrains the effects of stellar variability and improve\ch{s} power for detecting planets.

In principle, one could analyze the time series of successive spectra, but this is computationally challenging and may be difficult to interpret, so astronomers typically analyze low-dimensional time series composed of the measured RVs, additional summary statistics computed from the spectra and, when available, photometric observations.  
Potential summary statistics include classical activity indicators (e.g., excess emission in cores of Ca II H\&K lines), the shape of  spectral lines (e.g., FWHM \& BIS, see Fig. 4, \S3.2 of the \textbf{main text}), measurements of RVs and/or spectral line shapes from distinct regions of the spectrum (e.g., from different spectral orders, 
$\simeq~2$\AA chunks of the spectrum, or individuals lines), or the entire spectrum itself.  
In practice when using a large number of summary statistics, dimension reduction (e.g., Principal Components Analysis; PCA).  

As stellar activity is particularly problematic, most effort to date has focused on modeling magnetic activity.  
%
%
\citet{aigrain2012} argue that the RV effect of stellar spots could be predicted based on measurements of their photometric effect. \ch{They model the flux of light as a function of time $t$ as} $\Psi_0\{1-F(t)\}$, \ch{where $\Psi_0$ is a constant with dimensions of flux and $F(t)$ models the time variability of the photometric signal,  and model the  RV variation contamination as  $-\alpha \dot{F} F + \beta F(t)^2$ with $\alpha, \beta >0$ estimated from the light flux. 
For most targets, continuous photometric measurements are unavailable, as observational constraints often lead to to large gaps in time series.  
One can estimate $F$ and $\dot{F}$ by modeling $F$ as a Gaussian process (GP). 
GPs are stochastic processes $G(\vec t)$, function of a variable $\vec t$ such that for any $n$ values of $\vec t$, $\vec t_1,...,\vec t_n$, $(G(\vec t_1),...,G(\vec t_n))$ follows a multivariate Gaussian distribution. In general, $\vec t$ can be a vector, but for our purpose, we define it as the time $t$. }
This implies that the GP is defined by two quantities: its mean $m(t)$ and kernel, $k(t,t')$, equal to the covariance of $G(t)$ and $G(t')$ \citep{rasmussen2005, aigrain2022}. The posterior probability distribution of $G(t)$ conditioned on noisy measurements of $G$ at $ t_1,..., t_n$ can be computed efficiently from formulae 2.23-2.24 in \citet{rasmussen2005}. \ci{Its mean} provides an interpolator for measurements at the desired times. By modelling $F(t)$ as a Gaussian process, one can obtain an estimate of its derivative (and the corresponding uncertainty) at the times of spectroscopic observations. 

The above model can be considered as a special case of a multivariate Gaussian process latent variable model.  
Based on astrophysical considerations, \cite{rajpaul2015} postulated that the effect of stellar activity on measured RVs, and two activity indicators (BIS and $\log R'_{HK}$) could be be described by a latent Gaussian process $G(t)$,

\begin{align}
  \label{eq:RajpaulModel}
   & \Delta \mathrm{RV}(t) = V_c G(t) + V_r \dot{G}(t),\nonumber  \\
   & \Delta \mathrm{BIS}(t) = B_c G(t) + B_r \dot{G}(t),\nonumber \\
   & \Delta \log R'_{HK}(t) = L_c G(t).
\end{align}
where $V_c, V_r, B_c, B_r $ and $L_c$ are free parameters.

The process $G(t)$ is characterised by its mean function, chosen as null, and there are various choices of kernels parametrising the noise, as discussed in \S\ref{sec:stellar_variability_as_correlated_noise}. This framework, implemented in \cite{barragan2022}\footnote{ \url{https://github.com/oscaribv/pyaneti}}, has been further generalised by~\citet{jones2022}, \citet{gilbertson2020}\footnote{ \url{https://github.com/christiangil/GPLinearODEMaker.jl}} and \cite{delisle2022}\footnote{\url{https://gitlab.unige.ch/Jean-Baptiste.Delisle/spleaf}}, where the RV and arbitrary ancillary indicators are modelled as linear combinations of a Gaussian process and its first two derivatives.

In the framework of~\citet{rajpaul2015} and \citet{jones2022}, the expression for the likelihood is the same as Eq.~\eqref{eq:likelihood_simple} except that the data $\vec y$ is the concatenation of RV and the ancillary indicators.
The expression for  $\vec V$ involves the covariance of $G(t)$ and its derivatives, which can be expressed analytically~\citep{rajpaul2015, jones2022}.  With the modified likelihood, the Bayesian model comparison framework applies here also. It is also possible to model jointly the radial velocities computed in different spectral bands, and avail on the fact the amplitude of stellar activity effects changes with wavelength~\citep{reiners2010, cale2021}. As mentioned in \S \ref{sec:magnetic_activity},  \cite{luger2021_III} builds a Gaussian process model of the time series of spectra assuming that the dominant source of variability is related to stellar rotation. Supposing there are $n$ wavelength bins centered at wavelength $\lambda_i$, $i=1,...,m$ and $N$ measurements at times $t_j, j=1,...,N$. Instead of having a model for 3 time series as in \eqref{eq:RajpaulModel}, they have $m$ time series, 
the flux at $\lambda_i$ as a function of time $t_j$, $F(\lambda_i, t_j)$, and they compute the means and covariances of $F(\lambda_i, t_j)$ and $F(\lambda_{i'}, t_{j'})$ for all combinations of $i,j,i',j'$.  Code is available at \url{https://github.com/rodluger/starry_process}.
\citet{gilbertson2023} provide an alternative data-driven approach for modeling the flux at every pixel as a function of time  that simultaneously models variability in both stellar spectrum and atmospheric absorption, but 
does not assume stellar variabiltiy is rotationally modulated.
(Their code is avaliable at  \url{https://github.com/christiangil/StellarSpectraObservationFitting.jl 
}).
Most analyses in the literature assume a stationary noise model.  For observing campaigns spanning many years, it may be advantageous to allow for non-stationary GP kernels to account for changes due to long-term magnetic activity cycles. 
A general framework based on Gaussian process regression networks is provided in \citet{camacho2023}.





\subsection{Robustness to a change of model}

Although the physics of Doppler shift and planetary motion is precisely understood, the complex astrophysics responsible for stellar variability can only be approximated by a GP model.
If the likelihood or priors are poorly chosen, then the results may be unreliable. In particular, if correlated noise is ignored, then one can easily be overconfident about the statistical significance of a putative planet.  
For real planets, the mass and eccentricity estimators can be biased~\citep{zakamska2011,damasso2019, hara2019ecc}. 
In extreme cases, one can recognize remnant correlated noise by subtracting the best-fit planetary model (or better subtracting the predicted planetary signals while over planetary models) and looking for evidence of correlated  residuals. 
Whether there is correlation in the residuals can be assessed by computing the Bayesian evidence different kernels to describe them.  Another approach consists in computing periodograms (see \S\ref{sec:periodsearch})\citep[e.g.,][]{baluev2013_gj581} or variograms~\citep{hara2019ecc} from the residuals, or evaluating the predictive cababilities of the model \citep{hara2022}. \ci{We stress again the importance of testing the sensitivity of the results to priors and likelihood choices. }

As discussed in \S\ref{sec:bayesian_priors}, when working with datasets of fewer than $\sim$20 observations and low amplitude signals (e.g., close to the average noise level), it is important to perform a sensitivty test to check that the key conclusions are robust to different (but still reasonable) assumptions for the likelihood and prior.

 \section*{ACKNOWLEDGMENTS}
 
The authors thank Michaël Crétignier, Xavier Dumusque, Heather Cegla and François Bouchy for their helpful suggestions. 

\bibliographystyle{ar-style1.bst}
\bibliography{biblio.bib}

\end{document}